\documentclass[tighten,apj,twocolumn,numberedappendix]{emulateapj}

\usepackage{amsmath}
\usepackage{url}
\usepackage{xspace}
\usepackage[usenames,dvipsnames]{color}

\newcommand{\rmn}[1]{\mathrm{#1}}
\newcommand{\dd}{\rmn{d}}

\newcommand{\pig}{\pi^0-\gamma}
\newcommand{\p}{\rmn{p}}
\newcommand{\e}{\rmn{e}}
\newcommand{\gas}{\rmn{gas}}
\newcommand{\CR}{\rmn{CR}}

\newcommand{\mev}{\ensuremath{~\rmn{MeV}}}
\newcommand{\gev}{\ensuremath{~\rmn{GeV}}}
\newcommand{\kev}{\ensuremath{~\rmn{keV}}}
\newcommand{\gr}{$\gamma$-ray\xspace}
\newcommand{\grs}{$\gamma$ rays\xspace}
\newcommand{\Fexp}{\ensuremath{F_{\gamma,\rmn{exp}}}^{\CR}\xspace}
\newcommand{\ssigma}{\ensuremath{\sigma}\xspace}
\providecommand{\abs}[1]{\lvert#1\rvert}
\newcommand{\gal}{{\tt{gll\_iem\_v05}\xspace}}
\newcommand{\egal}{{\tt{iso\_clean\_v05}\xspace}}

\usepackage{xspace}
\def\Fermi{{\em Fermi}\xspace}
\renewcommand{\deg}{\ensuremath{^{\circ}}\xspace}
\slugcomment{accepted for publication in ApJ}

\shorttitle{Search for \gr\ Emission in Galaxy Clusters}
\shortauthors{Ackermann et al.}

\begin{document}


\title{Search for cosmic-ray induced gamma ray emission in Galaxy Clusters}
\author{The Fermi-LAT Collaboration: M.~Ackermann\altaffilmark{2{$\dagger$}} }
\author{M.~Ajello\altaffilmark{3} }
\author{A.~Albert\altaffilmark{4} }
\author{A.~Allafort\altaffilmark{5} }
\author{W.~B.~Atwood\altaffilmark{6} }
\author{L.~Baldini\altaffilmark{7} }
\author{J.~Ballet\altaffilmark{8} }
\author{G.~Barbiellini\altaffilmark{9,10} }
\author{D.~Bastieri\altaffilmark{11,12} }
\author{K.~Bechtol\altaffilmark{5} }

% ...
\altaffiltext{3}{Space Sciences Laboratory, 7 Gauss Way, University of California, Berkeley, CA 94720-7450, USA}
\altaffiltext{4}{Department of Physics, Center for Cosmology and Astro-Particle Physics, The Ohio State University, Columbus, OH 43210, USA}
\altaffiltext{5}{W. W. Hansen Experimental Physics Laboratory, Kavli Institute for Particle Astrophysics and Cosmology, Department of Physics and SLAC National Accelerator Laboratory, Stanford University, Stanford, CA 94305, USA}
\altaffiltext{6}{Santa Cruz Institute for Particle Physics, Department of Physics and Department of Astronomy and Astrophysics, University of California at Santa Cruz, Santa Cruz, CA 95064, USA}
\altaffiltext{7}{Universit\`a  di Pisa and Istituto Nazionale di Fisica Nucleare, Sezione di Pisa I-56127 Pisa, Italy}
\altaffiltext{8}{Laboratoire AIM, CEA-IRFU/CNRS/Universit\'e Paris Diderot, Service d'Astrophysique, CEA Saclay, 91191 Gif sur Yvette, France}
\altaffiltext{9}{Istituto Nazionale di Fisica Nucleare, Sezione di Trieste, I-34127 Trieste, Italy}
\altaffiltext{10}{Dipartimento di Fisica, Universit\`a di Trieste, I-34127 Trieste, Italy}
\altaffiltext{11}{Istituto Nazionale di Fisica Nucleare, Sezione di Padova, I-35131 Padova, Italy}
\altaffiltext{12}{Dipartimento di Fisica e Astronomia ``G. Galilei'', Universit\`a di Padova, I-35131 Padova, Italy}
% ...


\begin{abstract}
Current theories predict relativistic hadronic particle populations in clusters of galaxies in addition to the already observed relativistic leptons. In these scenarios hadronic interactions give rise to neutral pions which decay into \grs\, that are potentially observable with the Large Area Telescope (LAT) on board the \Fermi space telescope. We present a joint likelihood analysis searching for spatially extended \gr\ emission at the locations of 50 galaxy clusters in 4 years of \Fermi-LAT data under the assumption of the universal cosmic-ray model proposed by Pinzke \& Pfrommer (2010).
We find an excess at a significance of 2.7\,\ssigma which upon closer inspection is however correlated to \emph{individual} excess emission towards three galaxy clusters: Abell~400, Abell~1367 and Abell~3112. We discuss these cases in detail and conservatively attribute the emission to unmodeled background (for example, radio galaxies within the clusters).
Through the combined analysis of 50 clusters we exclude hadronic injection efficiencies in simple hadronic models above {21\%} and establish limits on the cosmic-ray to thermal pressure ratio within the virial radius, $R_{200}$, to be below {1.2-1.4\%} depending on the morphological classification. In addition we derive new limits on the \gr\ flux from individual clusters in our sample.
\end{abstract}


%\keywords{galaxy clusters: general -- gamma rays, cosmic rays, joint likelihood analysis}

\section{Introduction}
\label{sec:intro}
The quest for the first detection of high-energy \grs\ from galaxy clusters is still ongoing. While there have been \gr\ detections from radio galaxies in clusters such as NGC1275 \citep{1983ApJ...274..549S,2009ApJ...699...31A,2012A&A...539L...2A} and IC310 \citep{2010ApJ...723L.207A,2010A&A...519L...6N} in the Perseus cluster, as well as M87 in the Virgo cluster \citep{1994ApJ...426..105S,2009ApJ...699...31A,2003A&A...403L...1A}, no cluster-wide \gr\ emission has been detected so far. Previous reports of space-based cluster observations in the \gev\ band include \citealp{2003ApJ...588..155R,2010ApJ...717L..71A,2011arXiv1110.6863Z,2012JCAP...07..017A,2012MNRAS.427.1651H}. Ground-based observations in the energy band $\gtrsim100\gev$ were reported for Perseus and A2029 by \citet{2006ApJ...644..148P}, for A496 and A85 by \citet{2009A&A...495...27A}, for Coma by \citet{2009A&A...502..437A,2012ApJ...757..123A}, for A3667 and A4038 by \citet{cangaroo_clusters}, for Perseus by \citet{2010ApJ...723L.207A,2012A&A...541A..99A}, and for Fornax by \citet{2012ApJ...750..123A}. Nevertheless, current space-based \gr\ detectors such as the Large Area Telescope (LAT) on board the \Fermi satellite \citep{Atwood2009} may be able to detect \grs\ from galaxy clusters during its lifetime. 

The discovery and characterization of cosmic-ray (CR) induced \grs\ from clusters could not only serve as a crucial discriminator between different models for the observed cluster-wide radio emission \citep{2004A&A...413...17P,2012MNRAS.426..956B}, but could also be a signpost of the physical heating processes underlying feedback by active galactic nuclei (AGN) that has been proposed to solve the ``cluster cooling flow problem'' \citep{1991ApJ...377..392L, 2008MNRAS.384..251G, 2011A&A...527A..99E,2012ApJ...746...53F, 2013arXiv1303.4746W, 2013arXiv1303.5443P}. Moreover, any observed \gr\ emission from clusters necessarily requires a detailed understanding of the astrophysical \gr\ contribution before putting forward constraints on more exotic scenarios such as annihilation or decay signals from dark matter (DM) \citep{2010JCAP...05..025A,Dugger2010,Pinzke2011,2012JCAP...01..042H}. 

The intracluster medium (ICM) consists of a hot (1-10\kev) plasma, which has primarily been heated through collisionless shocks that form as a result of the hierarchical build up of galaxy clusters \citep[e.g.,][]{2003ApJ...593..599R,2006MNRAS.367..113P,2008ApJ...689.1063S,2009MNRAS.395.1333V}. These structure formation shocks and the turbulent motions of the gas in combination with intracluster magnetic fields provide the necessary conditions for efficient particle acceleration \citep[e.g.,][]{1998APh.....9..227C,2003ApJ...593..599R}. Recent radio observations prove the existence of relativistic electrons and magnetic fields in clusters. Some cool core (CC) clusters host radio mini halos in their centers \citep{2011A&A...527A..99E}. Additionally, a subsample of merging non-cool core (NCC) clusters show radio relics at their periphery \citep{2004rcfg.proc..335K} and/or giant radio halos  that often extend out to Mpc scales. While giant radio halos have been observed in more than 50 clusters \citep[see, e.g.,][for a review]{2008SSRv..134...93F}, their precise origin is still not understood.  
There are two competing theories to explain these radio halos. 

In \emph{hadronic models}, CR ions and protons (p) are accelerated in structure formation shocks, jets of radio galaxies, and supernovae-driven galactic winds \citep[see e.g.,][] {Volk1996,Ensslin1997,1997ApJ...487..529B,2004A&A...413...17P} and significant populations of CR protons can accumulate due to their long cooling time in the ICM \citep{Volk1996}. Inelastic collisions of CR ions with thermal protons of the ICM produce both neutral and charged pions, which decay almost instantly into \grs and electrons/positrons, respectively. This process could in principle account for the radio-emitting leptons, while requiring only a modest CR-to-thermal pressure ratio of (at most) a few percent \citep{1980ApJ...239L..93D,1982AJ.....87.1266V, 1999APh....12..169B, 2000A&A...362..151D, 2001ApJ...562..233M,2001ApJ...559...59M, 2003MNRAS.342.1009M,2003A&A...407L..73P, 2004A&A...413...17P,  2004MNRAS.352...76P, 2007IJMPA..22..681B, 2008MNRAS.385.1211P, 2008MNRAS.385.1242P, 2009JCAP...09..024K, 2010MNRAS.401...47D,  2010arXiv1003.0336D,2010arXiv1003.1133K, 2010arXiv1011.0729K,  2011A&A...527A..99E}. The non-detection of \gr\ emission from individual radio halo clusters places strong limits on intracluster magnetic fields within the hadronic model \citep{2004MNRAS.352...76P,2011ApJ...728...53J, 2012ApJ...757..123A, 2012MNRAS.426..956B, 2012A&A...541A..99A}. The secondary leptons also Compton upscatter with background radiation fields to \gr\ energies, but this emission is always subdominant compared to the \grs\ produced by decaying neutral pions \citep{2003MNRAS.342.1009M,2008MNRAS.385.1211P,Pinzke2010}. 

The \emph{re-acceleration models} assume the existence of a long-lived pool of mildly relativistic electrons, that were accelerated in the past by structure formation shocks, galactic winds, and AGN, or coincide with secondary electrons that are injected in the aforementioned hadronic CR p-p interactions. Those CR electrons scatter with plasma waves that are excited by ICM turbulence, e.g., after a cluster merger. These particle-wave interactions may accelerate the particles through the second order Fermi process to sufficiently high energies to explain the observed radio emission \citep{1987A&A...182...21S,2001MNRAS.320..365B,Petrosian2001,2004MNRAS.350.1174B, 2005MNRAS.363.1173B, 2007MNRAS.378..245B, 2011MNRAS.410..127B, 2009A&A...507..661B, 2013MNRAS.429.3564D}. 

Assuming that the same physical processes that produce \grs\ are present in each galaxy cluster, independent of mass, age and other characteristics, we employ a joint likelihood analysis to search for these \grs. The resulting universal scaling factor $A_\gamma$ can be used to derive limits on the hadronic acceleration efficiency at structure formation shocks, and the volume-averaged CR-to-thermal pressure $\langle X_\CR\rangle$. While the joint likelihood method can be applied to study any emission governed by a universal physical process, in this paper we focus on the search for \grs\ from CR-induced pion decay and defer more exotic scenarios such as \grs\ from dark matter annihilation and decay to future studies. 

The outline of this paper is as follows. We discuss our cluster selection in Section~\ref{sec:selection} and address the \Fermi-LAT observations and data analysis in Section~\ref{sec:analysis}. Our cluster emission models are described in Section~\ref{sec:signal_bg}. We present and discuss our results in Section~\ref{sec:results}. Finally, we present a survey of possible systematics in Section~\ref{sec:systematics}, and conclude in Section ~\ref{sec:summary}. Throughout the paper we assume a cosmology with $\Omega_{\rm{m}} = 0.3$, $\Omega_{\Lambda} = 0.7$ and $h=0.7$.

\section{Cluster Selection}
\label{sec:selection}
{Assuming a correlation between \gr\ and X-ray luminosity in galaxy clusters \citep{1998APh.....9..227C,2004A&A...413...17P}}, we use the extended HIghest X-ray FLUx Galaxy Cluster Sample (HIFLUGCS) \citep{hiflucs2001,chen2007}, containing the 106 nearby brightest X-ray clusters from the ROSAT all-sky survey, as a suitable list of sources when constructing our analysis sample.%%\ORdel{in what regard are HIFLUGCS clusters suitable for this study?}

Given that our statistical approach assumes independent sources, we remove clusters where the angular separation between cluster centers is less than any of the respective virial radii enlarged by 1\deg.\footnote{All cluster positions, which were taken from the NASA Extragalactic Database (\url{http://ned.ipac.caltech.edu/}), are based on observations in the optical waveband.} 
This cut is motivated by Monte Carlo (MC) studies (see Appendix~\ref{sec:sim_emin} for details),
showing that for an energy threshold of 500\mev, the bias on the likelihood ratio test statistic due to overlaps is minimal under the condition above. However, in the case that the expected \gr\ flux from a single HIFLUGCS cluster within such an ensemble of overlapping clusters is responsible for more than 90\% of the total expected emission, calculated using the approach in \citet{Pinzke2011}, we neglect the other clusters in the ensemble and attribute all photons to the cluster with the largest expected flux.

The aforementioned virial radius, $R_{200}$, of the cluster is the radius containing the virial mass, $M_{200}$, which in turn is derived from the $M_{500}$ mass reported by \citet{chen2007}.\footnote{We define the virial radius of a cluster as the radius at which the mean interior density equals 200 times the {\em critical density} of the universe.} We solve for $M_{200}$ using 
$M_{200}= M_{200}\times200/500\times[c_{200}(M_{200})/c_{500}(M_{200})]^3$ \citep{2005RvMP...77..207V}, where the halo-mass-dependent concentration parameter $c$ is derived from a power-law fit to a sample of observed galaxy cluster concentration and masses \citep{2007MNRAS.379..190C}.

The extended HIFLUGCS catalog contains clusters up to redshifts of 0.18, with the majority located  at $z<0.1$. For simplicity, we do not apply a redshift correction to the spectrum, and decided to exclude all clusters above z=0.1, as their inclusion without applying a correction would amount to incorrect modeling of the spectrum. These clusters contribute less than 1\% to the total expected \gr flux including all clusters in the HIFLUGCS catalog, as derived in \citet{Pinzke2011}, making this a suitable approach. In addition we exclude clusters that lie in a box defined by $\abs{b}\leq50\deg$ and $\abs{l}\leq20\deg$, as this region contains emission from the Galactic plane and the \Fermi bubbles \citep{SuFinkbeiner2010}, the models for which have relatively large systematic uncertainties. The Virgo cluster, which has a virial radius of $\sim\,8\deg$ on the sky \citep{2001A&A...375..770F} and clusters that fall into this region, such as M49, are excluded from the analysis.

The Galactic plane presents a substantial challenge due to a large number of potentially unresolved point sources and uncertainties in modeling the diffuse foregrounds. We conservatively define the plane region to be $|b|\leq20\deg$ and remove clusters from our sample that lie within this region, such as the Centaurus cluster. There are 9 Abell clusters present in the HIFLUCGS catalog that are located within a radius of 7\deg from the center of the Centaurus cluster which overlap with one another. In order to avoid highly crowded analysis regions, we exclude the entire region.

The remaining clusters are considered separately according to their classification as either CC, NCC or unclassified. For this purpose we use the classification presented in \citet{2010A&A...513A..37H} or follow recommendations by \citet{2009ApJS..182...12C}. Among other quantities, we classify clusters that have central entropy values $K_0\leq30\kev\mathrm{cm}^2$ as CC and otherwise as NCC. When there is no supporting X-ray data available, we leave these clusters unclassified.

\begin{figure}[tbp]
%\epsscale{.95}
\plotone{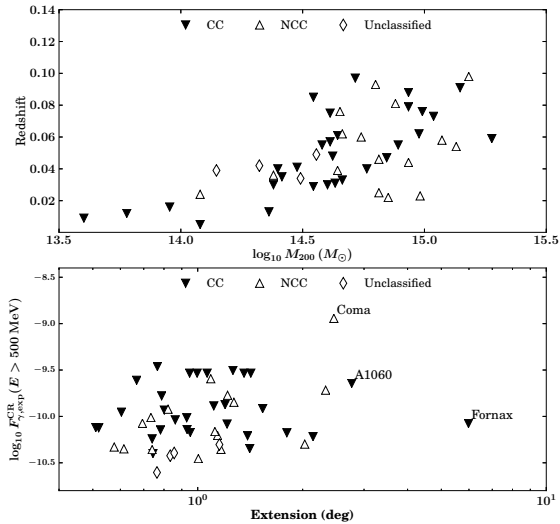}
\caption{\label{fig:clustersample} Summary of cluster quantities in our analysis. In the upper panel we show redshift versus $M_{200}$ and in the lower panel the predicted \gr\ flux above 500\mev\ taken from \citet{Pinzke2011}, $\Fexp(E>500\mev)$ versus its extension ($=2\times \theta_{200}$) in degrees. Most of the selected clusters in the sample have extensions of $\sim1\deg$.} 
\end{figure}

\begin{figure*}[tbp]
%\epsscale{.80}
%\plottwo{AllSky.eps}{AllSky.eps}
\plotone{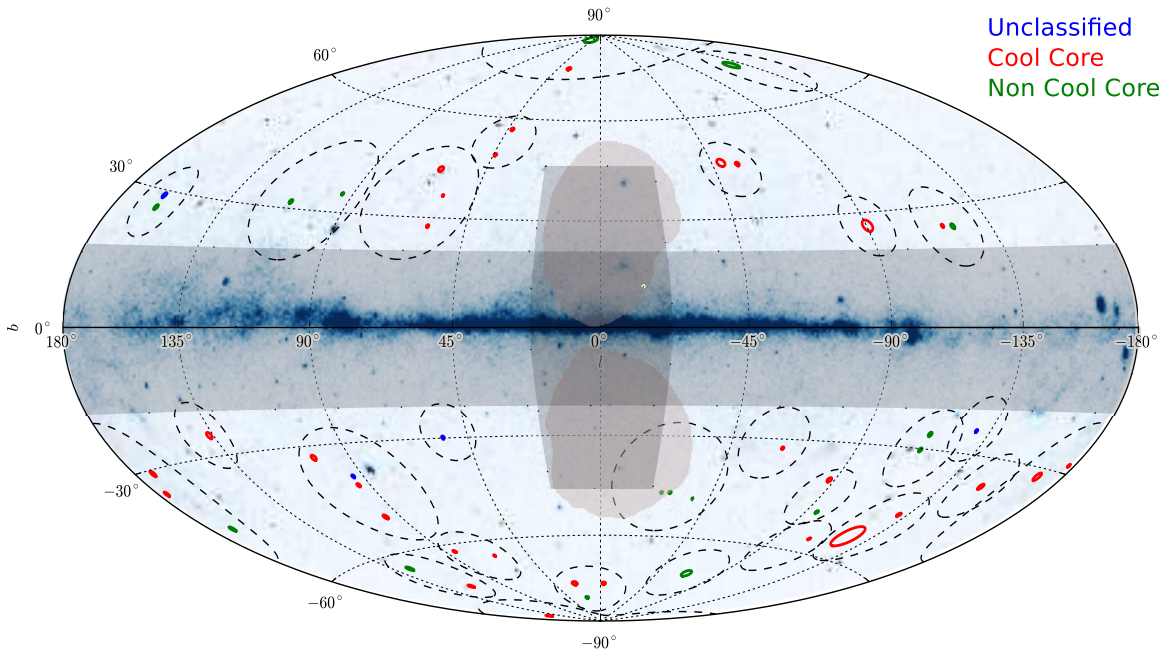}
\caption{(color online)\label{fig:allsky} Hammer-Aitoff projection of the sky as seen by the LAT after 4 years of exposure. Shown is a counts map generated for {\tt{CLEAN}} class events in the energy range from 500\mev\ to 200\gev\ using the same set of quality cuts as described in Section~\ref{sec:analysis}. The dashed circles correspond to the analysis regions considered for this analysis. The solid circles represent the clusters used in this analysis and their extension as characterized by their virial radii. In red we show CC, in green NCC, and in blue unclassified clusters. We shade the region which is described by the Fermi bubbles in \citet{SuFinkbeiner2010} whilst schematically overlaying our geometric cuts for masking the Galactic plane ($\abs{b}\leq20\deg$) and the Galactic bubbles ($\abs{b}\leq50\deg$ and $\abs{l}\leq20\deg$). The latter has been designed to mask out the majority of the emission that can be attributed to the lobes. When comparing our mask with the geometric description by \citet{SuFinkbeiner2010}, we found that 2 clusters are contained inside the bubbles. We checked that the region containing these two clusters is modeled well and hence keep the clusters in our analysis.}
\end{figure*}

The 50 galaxy clusters we consider in this analysis are listed in Table \ref{tab:cluster_overview_all} along with characteristic cluster quantities (see Figure~\ref{fig:clustersample}) and their classification. We show their location and radial extension on the sky in Figure~\ref{fig:allsky}.\footnote{\label{footnote:virialRadius}The extension we use is twice the angle subtended by the angular virial radius, $\theta_{200}$, which is given by $\theta_{200} = \arctan(R_{200}/D_{a})\times 180^\circ/\pi$, where $D_a$ is the angular diameter distance from the Earth to the center of the cluster.}

\begin{deluxetable*}{lrrrrrrrrrrr}
%\begin{deluxetable}{lrrrrrrrrrrr}
%\tabletypesize{\scriptsize}
%\rotate
%\tablenum{1}
\tablecaption{Cluster Sample considered for the Analysis}
\tablewidth{0pt}
\tablehead{
%% \colhead{Name} & \colhead{R.A.} & \colhead{Decl.} & \colhead{z} & \colhead{$R_{200}$} & \colhead{$M_{200}$} & \colhead{Extension} & \colhead{$\Fexp(E>500\mev)$} & $\langle X_{\CR}(R<R_{HL})\rangle$ & $\langle X_{\CR}(R<R_{200})\rangle$ & \colhead{Morphology}\\
%% & \colhead{(\deg)} & \colhead{(\deg)} & & \colhead{(Mpc)} & \colhead{($\times10^{14} M_\sun/H_{70}$)} & \colhead{(\deg)} & \colhead{$(10^{-10}\mathrm{ph\,cm^{-2}\,s^{-1}})$} & \colhead{($\times10^{-1}$)}& \colhead{($\times10^{-1}$)}&
%% }
\colhead{Name} & \colhead{R.A.} & \colhead{Decl.} & \colhead{z} & \colhead{$R_{200}$} & \colhead{$M_{200}$} & \colhead{Ext.} & \colhead{$\Fexp(E>500\mev)$} & $\langle X_{\CR}\rangle_{R_{HL}}$ & $\langle X_{\CR}\rangle_{R_{200}}$ & \colhead{Morph.}\\
& \colhead{(\deg)} & \colhead{(\deg)} & & \colhead{(Mpc)} & \colhead{($\times10^{15}$)} & \colhead{(\deg)} & \colhead{$(10^{-10}\mathrm{ph\,cm^{-2}\,s^{-1}})$} & \colhead{($\times10^{-1}$)}& \colhead{($\times10^{-1}$)}&
}
\startdata
2A0335  & 54.65 & 9.97 & 0.04 & 1.31 & 0.26 & 1.06 & 2.84 & 0.32 & 0.14 &  CC  \\
A0085  & 10.41 & -9.34 & 0.06 & 1.89 & 0.78 & 1.00 & 2.93 & 0.24 & 0.10 &  CC  \\
A0119  & 14.09 & -1.26 & 0.04 & 1.96 & 0.86 & 1.27 & 1.42 & 0.24 & 0.14 &  NCC  \\
A0133  & 15.66 & -21.96 & 0.06 & 1.52 & 0.41 & 0.78 & 0.72 & 0.28 & 0.13 &  CC  \\
A0262  & 28.21 & 36.15 & 0.02 & 0.91 & 0.09 & 1.54 & 1.21 & 0.41 & 0.30 &  CC  \\
A0400  & 44.41 & 6.03 & 0.02 & 1.02 & 0.12 & 1.17 & 0.44 & 0.38 & 0.28 &  NCC  \\
A0478  & 63.34 & 10.48 & 0.09 & 1.95 & 0.86 & 0.67 & 2.45 & 0.24 & 0.09 &  CC  \\
A0496  & 68.41 & -13.25 & 0.03 & 1.58 & 0.46 & 1.36 & 2.83 & 0.28 & 0.12 &  CC  \\
A0548e  & 87.16 & -25.47 & 0.04 & 1.06 & 0.14 & 0.76 & 0.25 & 0.37 & 0.26 &  ?  \\
A0576  & 110.35 & 55.74 & 0.04 & 1.56 & 0.44 & 1.14 & 0.63 & 0.28 & 0.16 &  NCC\tablenotemark{b}  \\
A0754  & 137.21 & -9.64 & 0.05 & 2.27 & 1.35 & 1.22 & 1.69 & 0.21 & 0.08 &  NCC  \\
A1060 (Hydra)  & 159.21 & -27.53 & 0.01 & 1.26 & 0.23 & 2.77 & 2.28 & 0.33 & 0.18 &  CC  \\
A1367 (Leo)  & 176.12 & 19.84 & 0.02 & 1.83 & 0.71 & 2.33 & 1.92 & 0.25 & 0.13 &  NCC  \\
A1644  & 194.31 & -17.35 & 0.05 & 1.83 & 0.70 & 1.11 & 1.30 & 0.25 & 0.14 &  CC  \\
A1795  & 207.25 & 26.59 & 0.06 & 2.02 & 0.95 & 0.95 & 3.01 & 0.23 & 0.11 &  CC  \\
A2065  & 230.68 & 27.72 & 0.07 & 2.11 & 1.09 & 0.86 & 0.92 & 0.21 & 0.08 &  CC  \\
A2142  & 239.57 & 27.22 & 0.09 & 2.30 & 1.40 & 0.77 & 3.45 & 0.29 & 0.14 &  CC  \\
A2199  & 247.16 & 39.55 & 0.03 & 1.52 & 0.40 & 1.42 & 3.00 & 0.27 & 0.13 &  CC  \\
A2244  & 255.68 & 34.05 & 0.10 & 1.66 & 0.52 & 0.52 & 0.76 & 0.25 & 0.14 &  CC  \\
A2255  & 258.13 & 64.09 & 0.08 & 1.87 & 0.76 & 0.69 & 0.85 & 0.22 & 0.11 &  NCC\tablenotemark{a}  \\
A2256  & 255.93 & 78.72 & 0.06 & 2.17 & 1.18 & 1.09 & 2.55 & 0.31 & 0.17 &  NCC\tablenotemark{a}  \\
A2589  & 351.00 & 16.82 & 0.04 & 1.38 & 0.30 & 0.95 & 0.68 & 0.30 & 0.13 &  CC  \\
A2597  & 351.33 & -12.11 & 0.09 & 1.45 & 0.35 & 0.51 & 0.76 & 0.28 & 0.17 &  CC  \\
A2634  & 354.58 & 27.03 & 0.03 & 1.55 & 0.43 & 1.39 & 0.62 & 0.26 & 0.13 &  CC  \\
A2657  & 356.21 & 9.14 & 0.04 & 1.71 & 0.58 & 1.21 & 0.83 & 0.28 & 0.15 &  CC  \\
A2734  & 2.83 & -28.87 & 0.06 & 1.58 & 0.46 & 0.74 & 0.44 & 0.26 & 0.12 &  NCC  \\
A2877  & 17.46 & -45.90 & 0.03 & 1.78 & 0.65 & 2.03 & 0.51 & 0.28 & 0.12 &  NCC\tablenotemark{a}  \\
A3112  & 49.47 & -44.24 & 0.08 & 1.53 & 0.41 & 0.60 & 1.11 & 0.27 & 0.14 &  CC  \\
A3158  & 55.67 & -53.63 & 0.06 & 1.68 & 0.55 & 0.82 & 1.20 & 0.20 & 0.09 &  NCC  \\
A3266  & 67.80 & -61.41 & 0.06 & 2.54 & 1.89 & 1.26 & 3.13 & 0.26 & 0.15 &  CC  \\
A3376  & 90.18 & -40.05 & 0.05 & 1.78 & 0.65 & 1.12 & 0.69 & 0.27 & 0.17 &  NCC  \\
A3822  & 328.53 & -57.85 & 0.08 & 1.57 & 0.45 & 0.61 & 0.45 & 0.28 & 0.17 &  NCC\tablenotemark{b}  \\
A3827  & 330.45 & -59.95 & 0.10 & 2.36 & 1.52 & 0.73 & 0.98 & 0.21 & 0.09 &  NCC\tablenotemark{b}  \\
A3921  & 342.41 & -64.39 & 0.09 & 1.77 & 0.63 & 0.58 & 0.47 & 0.26 & 0.13 &  NCC\tablenotemark{b}  \\
A4038  & 356.90 & -28.13 & 0.03 & 1.28 & 0.24 & 1.20 & 1.34 & 0.32 & 0.17 &  CC  \\
A4059  & 359.17 & -34.67 & 0.05 & 1.54 & 0.42 & 0.93 & 0.97 & 0.28 & 0.13 &  CC  \\
COMA  & 194.95 & 27.98 & 0.02 & 2.02 & 0.96 & 2.46 & 11.40 & 0.23 & 0.12 &  NCC  \\
EXO0422  & 62.93 & -29.81 & 0.04 & 1.30 & 0.25 & 0.93 & 0.72 & 0.32 & 0.17 &  CC  \\
FORNAX  & 54.63 & -35.46 & 0.01 & 1.01 & 0.12 & 5.98 & 0.84 & 0.38 & 0.25 &  CC  \\
HCG94  & 349.32 & 18.72 & 0.04 & 1.22 & 0.21 & 0.83 & 0.38 & 0.33 & 0.21 &  ?  \\
HYDRA-A  & 139.52 & -12.10 & 0.06 & 1.50 & 0.38 & 0.79 & 1.67 & 0.29 & 0.12 &  CC  \\
IIIZw54  & 55.32 & 15.40 & 0.03 & 1.45 & 0.35 & 1.41 & 0.45 & 0.30 & 0.16 &  CC  \\
IIZw108  & 318.48 & 2.57 & 0.05 & 1.47 & 0.36 & 0.86 & 0.40 & 0.29 & 0.19 &  ?  \\
NGC1550  & 64.91 & 2.41 & 0.01 & 0.81 & 0.06 & 1.80 & 0.67 & 0.45 & 0.33 &  CC  \\
NGC5044  & 198.85 & -16.39 & 0.01 & 0.73 & 0.04 & 2.14 & 0.61 & 0.50 & 0.42 &  CC  \\
RXJ2344  & 356.07 & -4.37 & 0.08 & 1.95 & 0.86 & 0.74 & 0.57 & 0.24 & 0.11 &  CC  \\
S405  & 57.89 & -82.22 & 0.06 & 1.56 & 0.44 & 0.74 & 0.40 & 0.28 & 0.18 &  CC\tablenotemark{b}  \\
S540  & 85.03 & -40.84 & 0.04 & 1.27 & 0.24 & 1.00 & 0.35 & 0.32 & 0.18 &  NCC  \\
UGC03957  & 115.24 & 55.43 & 0.03 & 1.39 & 0.31 & 1.15 & 0.50 & 0.30 & 0.15 &  ?  \\
ZwCl1742  & 266.06 & 32.98 & 0.08 & 2.04 & 0.98 & 0.80 & 1.17 & 0.23 & 0.10 &  CC  \\
\enddata
\tablecomments{The sample we used to carry out our analysis, locations taken from NED database. The columns from left to right read: cluster name according to \citep{hiflucs2001}, right ascension (J2000), declination (J2000), redshift, $R_{200}$, $M_{200}$ in units of $M_\sun/H_{70}$, extension ($2\times \theta_{200}$, where $\theta_{200}$ refers to the angular virial radius as described in footnote~\ref{footnote:virialRadius}), expected \gr\ flux above 500\mev, CR-to-thermal pressure ratio within $R_{200}$ and the half-light radius, $R_{HL}$ (see Section~\ref{sec:cosmic-rays} and footnote~\ref{footnote:rHL} for details), and morphological classification. We denote unclassified clusters with a ?. We note that columns 8-10 are based on the predictions in \citet{Pinzke2011} and assume $A_{\gamma}=1$. Redshifts from \citet{chen2007} and references therein. $M_{200}$ and $R_{200}$ are calculated from $R_{500}$ and $M_{500}$ in the aforementioned reference. Unless specified otherwise, all classifications are from \citet{2010A&A...513A..37H}.}
\tablenotetext{a}{Values and classifications taken from ACCEPT database \citep{2009ApJS..182...12C}}
\tablenotetext{b}{Values and classifications from \citet{2009ApJ...691.1787S}}
\label{tab:cluster_overview_all}
\end{deluxetable*}
%\end{deluxetable}

%\onecolumn
%\twocolumn
\section{\Fermi-LAT Observations and Data Analysis}
\label{sec:analysis}

The \Fermi Large Area Telescope (LAT) is the main instrument on board the \Fermi Gamma-ray Space Telescope. Since launch in 2008, the LAT has surveyed the \gr\ sky in the energy range from 20\mev\ to $>300\gev$ with unprecedented sensitivity. For more details about the LAT, the reader is referred to \citet{Atwood2009} and to \citet{Instrumentpaper} for the on-orbit LAT performance.
We carry out a binned likelihood analysis of 48 months of \Fermi-LAT data (2008-08-04 -- 2012-08-04), which has been reprocessed to account for the time-dependent calorimeter response \citep{ReprocessingPaper} and refer to this data as {\tt{P7REP}}. We select events corresponding to the {\tt CLEAN} class, which consists of the events that have the highest probability of being \grs, and use the {\tt P7REP\_V15} instrument response functions (IRFs) provided in the software package {\tt Fermi Science Tools v9r32p5}.\footnote{Both the LAT data as well as the appropriate analysis tools are made available to the public by the \Fermi Science Support Center \url{http://fermi.gsfc.nasa.gov/ssc/data/}} We apply standard quality cuts to our data using {\tt gtmktime} and require {\tt{DATA\_QUAL==1 \&\& LAT\_CONFIG==1}} which refers to the configuration during nominal science operation. We require the magnitude of the rocking angle of the LAT to be $\leq52\deg$ and reject events above a zenith angle of 100\deg to greatly reduce contamination by Earth limb emission. We use {\tt gtbin} to bin the data in 0.1\deg spatial bins.\footnote{It should be noted that the binned likelihood analysis using the {\tt Fermi Science Tools} uses square shaped analysis regions. For simplicity we refer to our ROIs by the central coordinates and radii of the ROI circumcircles.} The spectra are binned in 18 logarithmically spaced bins from 500\mev\ to 200\gev. Above 200\gev, the number of expected events given the models and the number of detected events are both sufficiently low that the models are not well constrained, so we omit this energy range from our analysis. 

\subsection{Joint Likelihood}
\label{sec:llh}
The joint likelihood is a source stacking technique, which has been previously applied with LAT data in the search for dark matter \citep{dwarfpaper} and to study the extragalactic background light \citep{2012Sci...338.1190A}. In brief, if the goal is to constrain or estimate a single or a  set of parameters common to a source class, then backgrounds and individual properties of each source can be modeled individually and treated as nuisance parameters in source-specific likelihoods. The source-specific likelihoods can then be multiplied to yield a joint likelihood function that is used for inference on the common parameter of interest. The joint likelihood function for our case can be written as:
\begin{equation}\label{Eq:compLike}
\begin{split}
\mathcal{L}\left(A_\gamma|\mathcal{D}\right)=\prod_{i}\mathcal{L}_{i}\left(A_\gamma,b_{i}|\mathcal{D}_{i}\right),
\end{split}
\end{equation} 
where $A_\gamma$ is a (dimensionless) universal scale factor which serves as the parameter of interest and $\mathcal{D}$ refers to the photon data for all ROIs. The physical interpretation of the universal scale factor in terms of CR-induced \gr\ emission is discussed further in Section~\ref{sec:cosmic-rays}. The $b_{i}$ parameters correspond to the parameters describing the background components in the individual regions-of-interest (ROIs) and are treated as nuisance parameters in the likelihood evaluation. The $b_{i}$s include the normalizations of the isotropic and Galactic diffuse components. We denote the photon data for each ROI as $\mathcal{D}_{i}$. The index $i$ runs over the ROIs in the sample. 
Having constructed the likelihood function, we use the profile likelihood method \citep[e.g.,][]{Rolke2005} to obtain best-fit values and confidence intervals for the parameter of interest, $A_{\gamma}$. We constrain our parameter of interest and nuisance parameters to be positive. The joint likelihood function is implemented in the {\tt Fermi Science Tools} using the {\tt Composite2} package and profiling over the likelihood function is achieved by means of {\tt MINOS} which is part of the {\tt MINUIT} package \citep{James1975}. The 90 \% confidence interval, $\left[A^{LL}_\gamma, A^{UL}_\gamma\right]$ is defined as the values of $A_\gamma$ where the log-likelihood has changed by 2.71/2 with respect to its value at the maximum, $-2\Delta\log{\mathcal{L}}=2.71$ \citep{Bartlett1953,Rolke2005}. These intervals can be reinterpreted as upper limits at the 95 \% confidence level (C.L.), if the parameter is unconstrained in the fit, which we do if the lower limit $\leq0$.

For quantifying the significance of a potential excess we employ the common likelihood ratio approach \citep{NeymanPearson1928}:
\begin{equation}\label{eq:globalts}
\begin{split}
TS=-2\log\left(\frac{\mathcal{L}(A_{\gamma}=0,\hat{\hat{b}})}{\mathcal{L}(\hat{A}_\gamma,\hat{b})}\right),
\end{split}
\end{equation} 
where $\mathcal{L}(A_\gamma=0,\hat{\hat{b}})$ is the null hypothesis, i.e. it represents the likelihood evaluated at the best-fit $\hat{\hat{b}}$ under the background-only hypothesis (Section~\ref{sec:nullfit}) and $\mathcal{L}(\hat{A}_\gamma,\hat{b})$ is the likelihood evaluated at the best-fit value of ${\hat{A}}_{\gamma}$, $\hat{b}$, when including our candidate $\gamma$-ray (cluster) source.

As we constrain the signal fit parameter $A_\gamma\geq0$, the null distribution of the test statistic, $TS$, is given by $\case{1}{2}\delta+\case{1}{2}\chi^2$ for one degree of freedom \citep{Chernoff1952}, which we verified by MC simulations. For details the reader is referred to Appendix~\ref{sec:validationsimulations}. While these simulations agree well with the expectations from Chernoff's theorem, studies based on random ROIs that encapsulate systematic effects in the LAT data (e.g. imperfect diffuse modeling, unresolved background sources, and percent-level inconsistencies in the IRFs) indicate that the significance estimated by simulations is probably somewhat too high when compared to the asymptotic expectations \citep[in prep.]{Dwarfs2}.

Recalling the discussion in Section~\ref{sec:selection}, sources and ROIs have to be selected such that the overlap is minimal since the joint likelihood function Eq.~\eqref{Eq:compLike} does not account for correlation terms between the different individual likelihood functions. For technical reasons, we cannot define a single ROI containing all 50 clusters, as this leads to an overflow in the number of free parameters that {\tt MINOS} is not able to handle. We therefore construct non-overlapping ROIs, each containing one or more (non-overlapping) cluster sources. 

\subsection{Construction of Regions of Interest}
\label{sec:regions}

In order to avoid overlaps between ROIs, we perform an iterative procedure in which we treat each cluster in our sample with its extension as listed in Table~\ref{tab:cluster_overview_all} as a seed source and construct a circular region around it. We require these regions to be at least 8\deg in radius in order to obtain a good fit to the background model. If we cannot accommodate clusters in separate ROIs we mitigate any remaining overlap by enlarging the ROIs such that more than one cluster may be contained in one analysis region. 

Using this method, we are able to construct 26 independent ROIs containing the clusters in our sample which are listed in Table~\ref{tab:roidefinition}. We show the locations and extensions of our selected clusters and the ROIs containing them in Figure~\ref{fig:allsky}. 

\begin{deluxetable}{lrrrrrr}
%% Keep a portrait orientation

%% Over-ride the default font size
%% Use 11pt
\tablewidth{0pt}
%\tabletypesize{\scriptsize}

%% Use \tablewidth{?pt} to over-ride the default table width.
%% If you are unhappy with the default look at the end of the
%% *.log file to see what the default was set at before adjusting
%% this value.

%% This is the title of the table.
\tablecaption{Regions of Interest considered in this Analysis}

%% This command over-rides LaTeX's natural table count
%% and replaces it with this number.  LaTeX will increment 
%% all other tables after this table based on this number

%% The \tablehead gives provides the column headers.  It
%% is currently set up so that the column labels are on the
%% top line and the units surrounded by ()s are in the 
%% bottom line.  You may add more header information by writing
%% another line between these lines. For each column that requries
%% extra information be sure to include a \colhead{text} command
%% and remember to end any extra lines with \\ and include the 
%% correct number of &s.
\tablehead{\colhead{Region} & \colhead{R.A.} & \colhead{Decl.} & \colhead{Radius} & \colhead{Clusters} & \colhead{CC} & \colhead{NCC}  \\ 
\colhead{} & \colhead{(\deg)} & \colhead{(\deg)} & \colhead{(\deg)} & \colhead{} & \colhead{} & \colhead{} } 

%% All data must appear between the \startdata and \enddata commands
\startdata
1 & 12 & $-$5 & 8 & 2 & 1 & 1 \\
2 & 15.66 & $-$21.95 & 8 & 1 & 1 & - \\
3 & 354 & $-$9 & 8 & 2 & 2 & - \\
4 & 172.07 & 21.02 & 8 & 1 & - & 1 \\
5 & 30.3 & 35.75 & 8 & 1 & 1 & - \\
6 & 257.77 & 73.43 & 16 & 2 & - & 2 \\
7 & 235 & 27.5 & 8 & 2 & 2 & - \\
8 & 259.5 & 40 & 16 & 3 & 3 & - \\
9 & 56.57 & 6.48 & 17 & 5 & 4 & 1 \\
10 & 58.2 & $-$31.75 & 10 & 2 & 2 & - \\
11 & 85.08 & $-$39.58 & 10 & 2 & - & 2 \\
12 & 43.67 & $-$85.28 & 10 & 1 & 1 & - \\
13 & 112.81 & 55.61 & 8 & 2 & - & 1 \\
14 & 138.36 & $-$10.87 & 10 & 2 & 1 & 1 \\
15 & 159.21 & $-$27.53 & 8 & 1 & 1 & - \\
16 & 321.97 & $-$60.77 & 16 & 3 & - & 3 \\
17 & 316.27 & 1.7 & 8 & 1 & - & - \\
18 & 44.13 & $-$45.61 & 8 & 1 & 1 & - \\
19 & 60 & $-$58 & 8 & 2 & 1 & 1 \\
20 & 195.95 & 28.37 & 14 & 2 & 1 & 1 \\
21 & 349.47 & 15.55 & 16 & 4 & 3 & - \\
22 & 197.2 & $-$18.49 & 8 & 2 & 2 & - \\
23 & 360 & $-$31 & 8 & 3 & 2 & 1 \\
24 & 17.46 & $-$45.9 & 8 & 1 & - & 1 \\
25 & 68.41 & $-$13.25 & 8 & 1 & 1 & - \\
26 & 87.51 & $-$25.09 & 8 & 1 & - & - \\
\hline Total  &    &    &    & 50 & 30 & 16 \\
\enddata

%% Include any \tablenotetext{key}{text}, \tablerefs{ref list},
%% or \tablecomments{text} between the \enddata and 
%% \end{deluxetable*} commands

%% General table comment marker
\tablecomments{We group neighboring clusters in non-overlapping analysis regions of interest (ROI) that are defined by center location and the associated radius. Note that with overlap we refer to the squared counts map inscribed in the circle defined with the coordinates in the table. We list the center of each region along with its radius together with its cluster content. Columns from left to right: ROI ID, R.A. (J2000), Decl. (J2000), radius of ROI, number of total clusters in ROI, number of CC-clusters in ROI, and number of NCC-clusters in ROI.}
\label{tab:roidefinition}
\end{deluxetable}

\section{Signal and Background Model}
\label{sec:signal_bg}

\subsection{Signal model: Gamma-Ray Emission from Cosmic Rays}
\label{sec:cosmic-rays}

In this paper we focus on the CR-induced \gr\ signal and defer an analysis of \grs\ originating from DM decay or annihilation to future studies. For a study of \grs\ originating from star-forming galaxies in galaxy clusters see, e.g., \citet{2012ApJ...755..117S}. Hydrodynamic simulations of galaxy clusters in a cosmological framework that include CR physics show an approximately universal spectral and spatial CR distribution within clusters as a result of hierarchical structure growth \citep{Pinzke2010}. The \gr\ emission induced by decaying neutral pions dominates over the inverse Compton emission from primary shock-accelerated electrons or secondaries injected in hadronic CR~p-p interactions within clusters \citep{2003MNRAS.342.1009M,2008MNRAS.385.1211P,Pinzke2010}.

Using a very simplified analytic model that employs a CR power-law energy
spectrum $dn_\rmn{CR}/d\varepsilon\propto \varepsilon^{-\alpha}$ with spectral
index $\alpha=2$ (i.e., equal CR energy density per logarithmic energy
interval), \citet{2009JCAP...08..002K} claim that the IC emission from primary
accelerated electrons at accretion shocks dominates over pion decay \grs by a
factor of $\simeq150 \,(\zeta_\e/\zeta_\p)\,(kT/10\, \rmn{keV})^{-1/2}$, where
$\zeta_\e$ and $\zeta_\p$ denote the fraction of shock-dissipated energy that is
deposited in CR electrons and protons, respectively. Instead of the centrally
concentrated pion decay emission, this model would give rise to a spatially
extended emission reaching out to the accretion shocks beyond the virial
radius. 

However, there are two simplifying model assumptions that conflict with results from numerical cluster simulations and observations of non-thermal emission {of clusters and supernova remnant shocks}, rendering the conclusions about this apparent dominance of the primary inverse Compton emission questionable. 

First, the assumed spectral index is in conflict with numerical
simulations of cosmological structure formation that have shown a continuous
distribution of Mach numbers with weaker flow and merger shocks being more
numerous in comparison to strong shocks with Mach numbers exceeding
$\mathcal{M}\gtrsim6$ \citep{2003ApJ...593..599R, 2006MNRAS.367..113P,
  2008ApJ...689.1063S, 2009MNRAS.395.1333V, 2011MNRAS.418..960V}.  This
necessarily implies a softer effective spectral injection index of primary
shock-accelerated particles of $\alpha_{\rmn{inj}}\simeq 2.3$, which is also
consistent with observed spectral indices of radio relics.{ In fact,
  elongated relics show a mean radio spectral index of
  $\langle\alpha_\nu\rangle=1.3$ \citep{2012A&ARv..20...54F}, which translates
  to a cooling-corrected spectral index of CR electrons of
  $\alpha=2\alpha_\nu=2.6$ (which is a lower limit since the uncorrected
  injection index could be as high as $\alpha=2\alpha_\nu+1=3.6$). Hence
  phenomenologically, the relevant shock strengths for radio relic emission are
  on average characterized by small Mach numbers of $\mathcal{M}\lesssim3$ and
  inconsistent with hard CR indices of $\alpha=2$.} It can be easily seen that
the different power-law indices are indeed the reason for the strongly differing
conclusions on the importance of primary inverse Compton emission. Electrons
with an energy of 500\gev\ can Compton up-scatter CMB photons to \gr\ energies
of around 1 GeV. Adopting the effective spectral injection index of primary
electrons, $\alpha_{\rmn{inj}}\simeq 2.3$, and assuming a post-shock temperature
at a typical accretion shock of $kT\simeq 0.1$~keV, we find a flux ratio of the
\citet{2009JCAP...08..002K} model and that by \citet{Pinzke2010} of $(500
\,\rmn{GeV}/0.1\,\rmn{keV})^{0.3\ldots0.6} \simeq 800\ldots6\times 10^5$.

% \sout{Second, the adopted ratio of injection efficiencies of
%   $\zeta_\e/\zeta_\p\sim 1$ are in conflict with successful models
%   for cluster radio relics (which yield for strong shocks
%   $\zeta_\e/\zeta_\p\sim 0.03$, Pinzke \& Pfrommer 2013). Moreover,}
Second, $\zeta_\e/\zeta_\p\sim 1$ is inconsistent with the observed
ratio for Galactic CRs, which has a differential energy ratio of
$K_\rmn{e/p}\simeq 0.01$ at 10~GeV \citep{2002cra..book.....S} {and
  observations of supernova remnants constrain the ratio
  $\zeta_\e/\zeta_\p\sim 0.001$ \citep{2011MNRAS.414.3521E,
    2012A&A...538A..81M}. Assuming universality of the shock
  acceleration process, statistical inferences about the CR
  distributions at the solar circle and non-thermal modeling of
  individual supernova remnants indicate that electron acceleration
  efficiencies are very subdominant in comparison to that of protons.}

%{\sout{Hence, we apply the simulation-based analytical approach for the CR distribution \citep{Pinzke2010}, which only requires the gas density profile inferred from X-ray measurements as input. Note that these simulations only account for advective CR transport, where the CRs may be tied to the cluster plasma via small-scale tangled magnetic fields. Additional CR transport processes such as CR diffusion and streaming have been neglected despite the possible importance  for explaining the radio-halo bimodality \citep{2011A&A...527A..99E,2012arXiv1207.6410Z,2013arXiv1303.4746W} or the physical heating process underlying ``feedback'' by AGN \citep{1991ApJ...377..392L, 2008MNRAS.384..251G, 2013arXiv1303.5443P}.}}
{Hence, as a baseline model we apply the simulation-based analytical approach for the CR distribution \citep{Pinzke2010}, which only requires the gas density profile inferred from X-ray measurements as input. Note that these simulations account only for advective CR transport, where the CRs may be tied to the cluster plasma via small-scale tangled magnetic fields with cored CR profiles as a consequence. Additional CR transport such as CR diffusion and streaming have been neglected.} % despite the possible importance for explaining the radio-halo bimodality \citep{2011A&A...527A..99E,2013arXiv1303.4746W,2013arXiv1311.4793Z} or the physical heating process underlying ``feedback'' by AGN \citep{1991ApJ...377..392L, 2008MNRAS.384..251G, 2013arXiv1303.5443P}. 
Furthermore we neglect the potentially important contribution from re-accelerated CRs \citep[e.g.][]{2007MNRAS.378..245B, 2011MNRAS.410..127B}. %{\sout{that typically produce flatter CR profiles and steeper CR spectra at high energies (similar to what we expect from the CR diffusion and streaming models).}} %These spectral and spatial CR features may be necessary to explain recent observations of the Coma radio halo \citep[e.g.,][]{2013MNRAS.429.3564D}. }

% In contrast, we find that the \gr emission from clusters is inconclusive regarding the preferred spatial CR profile while we postpone a detailed comparison of different CR spectra to future work. For more details on the impact of different spatial CR profiles see the discussion in Section~\ref{sec:scalefactor} and ~\ref{sec:upperLimits}.

The pion decay \gr\ flux above energy $E$ as a result of hadronic CR interactions, $\Fexp(>E)$, can be parametrized as:
\begin{equation}
\label{eq:CR_flux}
\Fexp(>E) = A_\gamma\,\lambda_{\pi^{0}\rightarrow \gamma}\left(>E\right) \int_V \dd V \kappa_{\pi^{0}\rightarrow \gamma}(R)\,,
\end{equation}
where the integral extends over the cluster volume $V$, $\lambda_{\pi^{0}\rightarrow \gamma}\left(>E\right)$ is the spectral \gr\ distribution and $\kappa_{\pi^{0}\rightarrow \gamma}(R)$ the spatial distribution, both given in \citet{Pinzke2010}. The parameter, $A_\gamma$, is a dimensionless universal scale factor, common to all the clusters in the (sub)sample. The predicted \gr\ flux above the minimum energy threshold 500\mev, $\Fexp(E>500\mev)$ is calculated using the formalism in \citet{Pinzke2011}. We tabulate these values in Table~\ref{tab:cluster_overview_all}. The quoted values of $\Fexp(E>500\mev)$ correspond to a maximal efficiency $\zeta_\rmn{p,max}=0.5$ for diffusive shock acceleration of CR ions at structure formation shocks which translate into $A_\gamma=1$ with correspondingly smaller values of $A_\gamma$ for smaller efficiencies (obeying however a non-linear relation). For completeness we show the CR formalism in Appendix~\ref{sec:CRs}.

The cluster brightness profile is used to fit the emission from each cluster and is derived from the line-of-sight integral of the \gr\ emissivity
\begin{eqnarray}
\nonumber S_\gamma(\psi,>E) &=&A_\gamma\,\lambda_{\pi^{0}\rightarrow \gamma}\left(>E\right)\\ \nonumber &\times&  \int_{\Delta\Omega} \dd\Omega \\
&\times& \int_\rmn{l.o.s} \dd l \,\frac{\kappa_{\pi^{0}\rightarrow \gamma}\left(R(l)\right)}{4\pi}\,.
\label{eq:CR_SB}
\end{eqnarray}
where $l$ is the line of sight (l.o.s.) distance in the
direction $\psi$ that the detector is pointing and $R(l)=\sqrt{l^2+D_a^2-2 D_a
l\cos\Psi}$ is the cluster radius. Here $D_a$ is the angular diameter distance from the Earth to the center of the cluster halo and $\cos\Psi\equiv\cos\theta\cos\psi-\cos\varphi\sin\theta\sin\psi$, with $\theta$ being the azimuthal and $\phi$ the polar angle, respectively. The angular integration $\dd \Omega= \sin\theta\dd \theta \,\dd \varphi$ is performed over a cone centered around $\psi$. The spatial features of our model are described in more detail in Appendix~\ref{sec:CRs}.

\begin{figure}[tbp]
\epsscale{.90}
\plotone{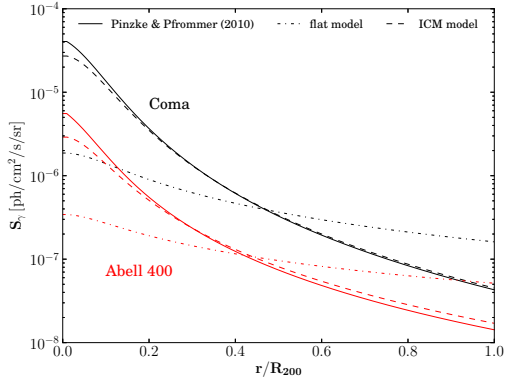}
\caption{{(color online)\label{fig:spatialModels} Expected surface brightness profiles for the three spatial models considered in this analysis. We show the profiles for two clusters, Coma (black lines) and Abell 400 (red lines), which have comparable distances ($z=0.02$) such that the flux difference corresponds to the difference in mass. We note that the ICM model (dashed line) only shows small differences with respect to our simulation-based baseline model (solid line). In contrast, the flat CR pressure model (dashed-dotted line) implies a flattening towards the outskirts of the cluster.}}
\end{figure}

{Outside the very center ($r > 0.03 R_{200}$), this model predicts a rather flat CR-to-thermal pressure profile, i.e., $\langle X_{\rmn{CR}} \rangle\sim\rmn{const}$. Most of the emission is contributed from the region around the core radius, which is well outside the central parts. In order to {{compare the chosen spatial CR profile, we contrast our analysis}} of the simulation-based model with two additional configurations in which the CR profile is derived from a constant $X_{\rmn{CR}}$ profile ({\it ICM model}) and a constant $P_\CR$ profile ({\it flat model}). The normalizations of these CR profiles are fixed by assuming that the total CR number within $R_{200}$ in our baseline model is conserved.\footnote{Note that CR streaming and diffusion---as spatial transport processes---conserve the total number of CRs. However, outward streaming changes their number {\em density} as a function of radius and transforms a peaked CR profile into an asymptotically flat one. Hence, to map an advection-dominated profile (i.e., our simulation-based baseline model) to the corresponding asymptotically flat profile that results from streaming, we compute the volume integral of the number density before and after CR streaming and normalize the latter such that the total number of CRs is conserved.} For illustration purposes, we show the surface brightness profiles using our three spatial emission models in Figure~\ref{fig:spatialModels} for the case of the (massive) Coma cluster and the much less massive cluster Abell~400.}

\begin{figure}[tbp]
%\epsscale{.90}
\plotone{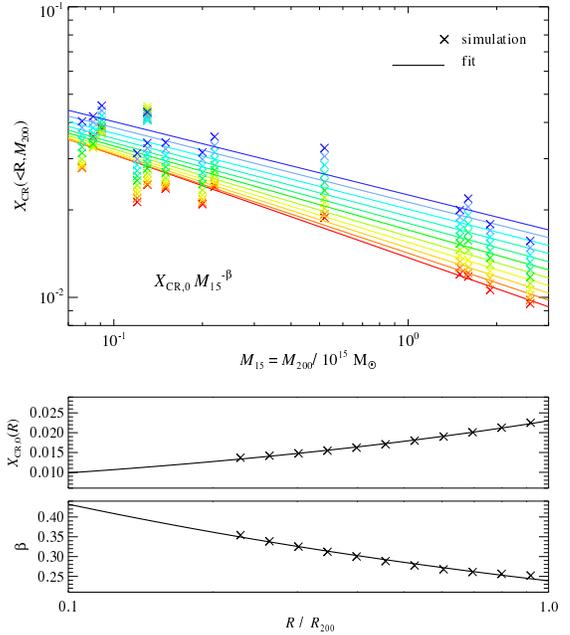}
%\plottwo{fit.XCR.eps}{fit.XCR.eps}
\caption{(color online)\label{fig:XCR_fit} Relative cosmic-ray pressure $X_\rmn{CR}$ within radius $R$. We show the ratio between cosmic-ray pressure and thermal pressure for 14 simulated galaxy clusters with different mass. In the upper panel the color scheme shows the relative pressure within different radii, in descending order from $1.0\times R_{200}$ to $0.2\times R_{200}$ (equally spaced in $\log R$). Each simulated cluster is denoted by an $\times$ and the mass dependence of $X_\rmn{CR}$ as a function of radius are denoted by the solid lines. The middle panel shows the radial dependence of the normalization of $X_\rmn{CR}$. The lower panel shows the radial dependence of the slope of the mass dependence of  $X_\rmn{CR}$. We find that the relative pressure increases as a function of radius, but decreases with increasing cluster mass.}
\end{figure}

In our framework the derivation of $A_{\gamma}$ also allows us to constrain the CR-to-thermal pressure ratio, $X_\CR$, in the ICM using the virial mass and virial radius of each cluster in our sample, since $A_\gamma \propto \langle X_\CR\rangle$, where $\langle X_\CR\rangle = {\left\langle P_\CR \right\rangle}_{V} / {\left\langle P_\rmn{th}\right\rangle}_{V}$ and the brackets indicate volume averages. To this end, we make use of the set of 14 galaxy cluster simulations presented in \citet{Pinzke2010}, which span almost two orders of magnitude in mass and include different dynamical states ranging from relaxed to merging clusters. We show the CR-to-thermal pressure ratio $X_\CR$ as a function of radius and cluster mass in Figure~\ref{fig:XCR_fit}. $X_\CR$ decreases for smaller radius approximately inversely with gas temperature since a composite of CRs and thermal gas favors the gas component over CRs upon adiabatic compression
\citep[e.g.,][]{2004A&A...413...17P}. 
At fixed radius, $X_\CR$ has a negative trend with mass, that is mainly driven by the virial temperature scaling of the
thermal pressure distribution. We have $\langle X_\CR\rangle \propto \langle C\rangle / \langle P_\rmn{th}\rangle \propto \tilde{C} / kT_{200}\sim M_{200}^{-0.23}$, where $C$ is the normalization of the CR distribution function and $\tilde{C} = C m_p/\rho$ denotes the dimensionless normalization, which scales with cluster mass as
$\langle\tilde{C}\rangle_V \propto M_{200}^{0.44}$ \citep{Pinzke2010} and partially offsets the virial mass scaling of the cluster temperature $kT_{200}\propto M_{200}^{2/3}$. 

To formalize these considerations, we fit an empirical relation of $X_\rmn{CR}(R/R_{200},M_{200})$ to the simulated data points and obtain
\begin{eqnarray}
\nonumber X_\rmn{CR} &=& 0.023 A_\gamma \left(\frac{R}{R_{200}}\right)^{0.369}\\
  &\times& \left(\frac{M_{200}}{10^{15}\,\rmn{M}_\odot}\right)^{-0.239
    \left(\frac{R}{R_{200}}\right)^{-0.258}}\,,
\end{eqnarray}
where $A_\gamma$ refers to our universal scale factor derived in the joint likelihood analysis. We tabulate the values for $\langle X_{\CR}\rangle$ for two different integration radii, $R_{200}$ and the half-light radius $R_{HL}$ assuming $A_{\gamma}=1$ in Table~\ref{tab:cluster_overview_all}.\footnote{\label{footnote:rHL} Here we define the half-light radius as the radius within which half of the emission originates. Note that this radius is usually smaller than $R_{200}/2$.}

\subsection{Background Model}
\label{sec:bkgModel}
The background model for each ROI in this analysis includes templates for the diffuse Galactic and isotropic emission components as well as individual \gr\ sources reported in the 2nd \Fermi-LAT catalog \citep[2FGL]{Nolan2012}. We model extended 2FGL sources according to the spatial templates provided by the \href{http://fermi.gsfc.nasa.gov/ssc/data/access/lat/2yr\_catalog/gll\_psc\_v07\_templates.tgz}{\Fermi Science Support Center}.
Unless stated otherwise we use the standard diffuse and extragalactic \gr background templates that are recommended for performing data analysis of reprocessed LAT data.\footnote{For the Galactic diffuse emission we use the template \gal\ and for the isotropic \gr\ background \egal.} We note that the Galactic diffuse emission model we use includes a residual component of diffuse \gr\ emission that is not modeled by any template. This component is smoothly varying and does not contribute importantly to the intensity $>500\,\mev$ in the regions considered here.\footnote{Further details on this new model can be found on the \href{http://fermi.gsfc.nasa.gov/ssc/data/access/lat/BackgroundModels.html}{FSSC website}.}

In the background model for each ROI, we include the union of 2FGL sources within the ROI radius enlarged by 5\deg and 2FGL sources located within 10\deg of any galaxy cluster in the ROI.

\subsection{Free Parameters of the Background Model}
Since the average virial radius of clusters in our sample is less than 2\deg, we leave the normalizations of all sources free that are contained within a 4\deg radius around each cluster. In addition, we allow the normalization of the templates used to model the Galactic foreground and isotropic diffuse emission to vary freely.

One shortcoming of using the 2FGL catalog (based on 2 years of LAT observations) to search within a dataset covering 4 years is that spectral parameters (in particular for variable sources) may have substantially changed. To account for this variability, we free the normalization of sources that coincide with bright spatial residuals, and determine their values through performing a fit using a background-only model to obtain the best-fit for the null-hypothesis. 

This procedure produces a large number of free parameters which are then fixed to their maximum likelihood values from the background-only model fit when maximizing the joint likelihood function introduced in Section~\ref{sec:llh} in order to avoid an overflow of free parameters and to ensure convergence of the maximization. The normalizations of the Galactic and isotropic diffuse components are left free in each ROI for the joint likelihood fit.

\section{Results and Discussion}
\label{sec:results}
We perform our analysis first by treating all 50 clusters in one common set. We call this the combined sample. Recalling the discussion in Section~\ref{sec:selection}, we then separately investigate CC and NCC clusters. 

\subsection{Background-only fit}
\begin{figure}[tbp]
%\epsscale{.9}
\plotone{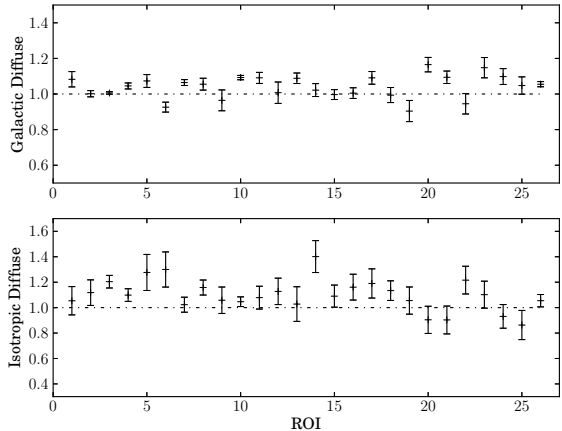}
\caption{\label{fig:diffuse_normalizations} Fit results for the normalization for the Galactic diffuse emission (top) and the isotropic diffuse emission (bottom) in each of the 26 analyzed ROIs. The dot-dashed lines indicate a nominal value of one associated to diffuse templates exactly modeling the emission in the regions. Our regions show a narrow scatter for the Galactic diffuse emission and a slightly larger scatter for the isotropic diffuse component. The latter is also associated with a minor bias towards normalization values $>1$.}
\end{figure}
\begin{figure}[tbp]
%\epsscale{.95}
\plotone{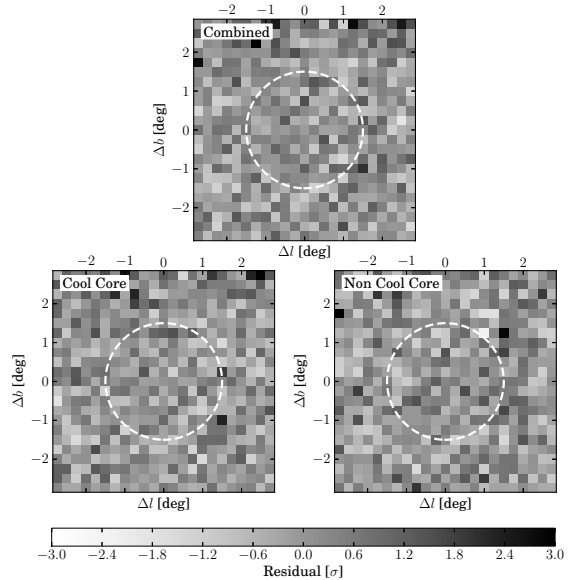}
\caption{\label{fig:datastack}Spatial residual significance maps for the combined sample (top) and the two sub-samples (bottom). The dashed circle corresponds to a radius of 1.5\deg which covers the majority of the emission region, assuming that the emission is contained within the cluster's virial radius. The stacked maps are created using the galaxy cluster centers listed in Table~\ref{tab:cluster_overview_all}. We do not observe any significant excess in the combined sample. If observed, any excess in these maps would be extended on at least the scale of the effective point spread function (PSF) of the LAT. Each pixel has a size of 0.25\deg.}
\end{figure}

\label{sec:nullfit}
Our ROIs are well described individually by the null hypothesis, i.e., despite the increase in data volume, our results are consistent with emission from only previously detected individual \Fermi-LAT sources and diffuse emission provided by the Galactic and extragalactic diffuse models respectively with the exception of three ROIs: the ROIs containing \object{Abell~400} and, less prominently \object{Abell~1367} and \object{Abell~3112} exhibit residual emission located within the virial radius of each respective cluster. We leave these excesses unmodeled in our baseline analysis and address the interpretation of these residuals in Section~\ref{sec:exceptions}. 
It is also reassuring that the fitted normalizations of the two global diffuse backgrounds are narrowly scattered around the nominal value of 1 for both the extragalactic (isotropic) and the Galactic diffuse component (refer to Figure~\ref{fig:diffuse_normalizations}) across our entire sample. The isotropic component shows a slight bias towards normalizations $>1$ which however has negligible effect on our results.
\noindent
Figure~\ref{fig:datastack} shows a stacked residual significance map for the full sample and CC and NCC sub-samples. These residual significance maps are created from theoretical model maps of predicted counts from the best-fit null hypothesis model summed over each cluster location. The combined model maps $M$ are then subtracted from the stacked counts maps $C$ from the same region  and the residual significance $R$ is computed as $R=(C-M)/\sqrt{M}$. For the spectral residuals we refer the reader to Appendix~\ref{sec:spectra}. No obvious excess is visible in either the spectral or spatial residuals. 
%\onecolumn
%\twocolumn

\subsection{Global Significance and Constraints on common Scale Factor $A_\gamma$}
\label{sec:scalefactor}
\begin{figure}[tbp]
%\epsscale{.95}
\plotone{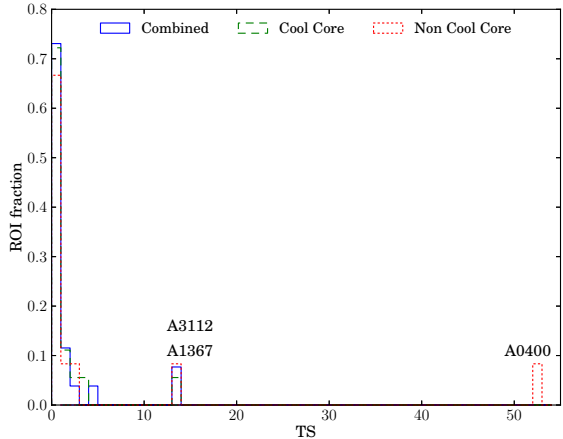}
\caption{(color online)\label{fig:ROI_TS} Distribution of ROI $TS$ values for full sample (blue, solid line), Cool Core (red, dashed line) and Non-Cool Core (green, dash-dotted line). We discuss the notable exceptions with $TS\geq9$ in Section~\ref{sec:exceptions}.}
\end{figure}

We then repeat the fitting procedure including a model of our galaxy clusters with the predicted \gr\ flux $F_\mathrm{exp}(E>500\mev)$ and using the spatial template and spectral form proposed by \citet{Pinzke2011}, leaving only $A_\gamma$ to vary freely. We show the distribution of associated $TS$ values for the respective samples in Figure~\ref{fig:ROI_TS}.
Assuming that backgrounds are properly modeled, and that CR physics governing the \gr\ emission of clusters is indeed universal, we calculate the best-fit value of the combined scale factor for the full sample along with the two morphological sub-samples. 
While the global $TS$ value of the scale factor for the full sample of 50 clusters is 7.3, corresponding to a formal 2.7\,\ssigma evidence, we note that the largest contributors to this tentative signal are from the aforementioned excess spatially coincident with \object{Abell~400} and also from less prominent excesses towards \object{Abell~1367} and \object{Abell~3112}. We discuss these special cases in the next section. Removing these clusters from the sample results in a drop of the significance below 2\,\ssigma yielding an upper limit to the common scale factor $A_\gamma^{UL}=0.29$ at 95\% C.L. for the whole cluster sample containing the remaining 47 galaxy clusters. While individually, the excess towards Abell~400 yields a higher significance than the excesses towards either Abell~3112 or Abell~1367, in the combined limit the contribution of this excess becomes negligible due to the lower flux prediction (which mainly determines the weight assigned to each cluster in the joint analysis), as compared to, e.g. the contribution from Coma that dominates the upper limit.

Since at present we can neither claim nor refute the origin of the observed excesses being due to \grs\ from the ICM, we calculate upper limits on the universal scaling factor, leaving those respective excesses unmodeled. For our whole sample we find a combined limit of $A^{UL}_\gamma=0.41$ at 95\% C.L. Considering the NCC/CC subsamples we find weaker limits $A^{UL}_\gamma=0.47$ on the scale factor for NCC systems as compared to $A^{UL}_\gamma=0.49$ for the CC subsample.

{We also calculated the limit on the combined scale factor for our two alternative spatial CR profiles. For the ICM model we obtain $A^{UL}_\gamma=0.48$ and for the flat model the combined limit is roughly a factor 4 larger with respect to the results from the baseline model, yielding $A^{UL}_\gamma=1.78$, at 95\% C.L. The associated global $TS$ values are 7.2 and 9.7 respectively.\footnote{Removing Abell~400, Abell~1367 and Abell~3112 from the cluster sample yield smaller global $TS$ values of 2.8 for the ICM model and 4.7 for the flat model, respectively.} %Interestingly the $TS$ values between the flat and ICM models in the CC cluster subsample show only minor differences while the values in the NCC cluster sample are a factor two larger for the flat model. This discrepancy, however, can easily be accounted for by the the large uncertainties in both the small $TS$ values and the normalization of the flat CR profiles. 
In addition a flat CR profile is preferred in the case of the Coma cluster which is also the cluster that contributes most to the constraints in the NCC cluster subsample. We provide the final values for our three setups in Table~\ref{tab:scalefactors}.} 

%% there are and how to align them.
\begin{deluxetable}{cccc}
\tablewidth{\columnwidth}
\tablecaption{Joint Scale Factor Limits}
%% The \tablehead gives provides the column headers.  It
%% is currently set up so that the column labels are on the
%% top line and the units surrounded by ()s are in the 
%% bottom line.  You may add more header information by writing
%% another line between these lines. For each column that requries
%% extra information be sure to include a \colhead{text} command
%% and remember to end any extra lines with \\ and include the 
%% correct number of &s.
\tablehead{\colhead{Model} & \colhead{Combined} & \colhead{CC} & \colhead{NCC} \\ 
\colhead{} & \colhead{} & \colhead{} & \colhead{} } 

%% All data must appear between the \startdata and \enddata commands
\startdata
Pinzke \& Pfrommer (2010) & 0.40 (7.4) & 0.49 (4.7) & 0.44 (2.4) \\
constant $X_\CR$ (ICM model) & 0.48 (7.2) & 0.64 (4.9) & 0.49 (2.4) \\
constant $P_\CR$ (flat model) & 1.78 (9.7) & 3.02 (5.2) & 1.71 (5.0) \\
\enddata

%% General table comment marker
\tablecomments{\label{tab:scalefactors}{Summary of joint scale factors. Columns 2-4 indicate the 95\% upper limit on the joint scale factor $A_\gamma$ for the respective samples. The global $TS$ values of each setup and sample are given in brackets.}}

\end{deluxetable}

For the combined sample we exclude $A_\gamma=1$ at more than 5\,\ssigma confidence, while for CC/NCC it is excluded at more than 4\,\ssigma confidence.
%\section{Discussion}
\subsection{The case of Abell~1367, Abell~3112 and Abell~400}
\label{sec:exceptions}
\begin{figure}
%\epsscale{.95}
\plotone{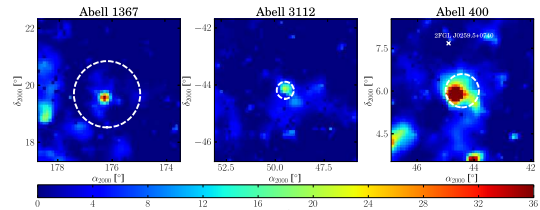}
\caption{(color online)\label{fig:tsmaps_clusters} $TS$ maps from an unbinned search in a $5\deg\times5\deg$ region centered on each of our notable clusters, Abell~1367, Abell~3112 and Abell~400. All excesses are found within the assumed cluster virial radius (dashed white circle), albeit marginally offset from the respective cluster centers (0.3\deg for both Abell~400 and Abell~1367 and 0.1\deg for Abell~3112). Each pixel has a width of 0.1\deg. The white $\times$ indicates the best-fit position of a previously detected 2FGL point source.}
\end{figure}

We obtained a $TS$ value of 13.8, corresponding to a pre-trial factor significance of 3.7\,\ssigma individually for each of the candidate \gr\ sources at the locations of Abell~1367 (Leo cluster) and Abell~3112. Conservatively assuming a binomial distribution for the trial probability, we find that these excesses correspond to a post-trial significance of 2.6\,\ssigma.
For each of these regions we calculated a $TS$ map using an unbinned likelihood method in a $5\deg\times5\deg$ region centered around each cluster with a grid spacing of 0.1\deg between test positions. The $TS$ value of a putative point-like source with a hadronic CR spectrum is evaluated at each test position on the grid to create a spatial map of the excess emission. We show these $TS$ maps in Figure~\ref{fig:tsmaps_clusters}. 
%\onecolumn
%\twocolumn
In both cases we find that the excess emission, albeit marginally offset from the center of the cluster (0.3\deg for \object{Abell~1367} and 0.1\deg for \object{Abell~3112}) is still contained within the virial radius of the respective cluster. For \object{Abell~1367} the difference in $TS$ obtained at the center of the cluster and the peak $TS$ position as shown in Figure~\ref{fig:tsmaps_clusters} is 15, while for \object{Abell~3112} the offset is smaller than the resolution used to create the $TS$ maps. In order to test whether the emission is more appropriately modeled assuming an extended emission profile over a new point source, we follow the analysis presented in \citet{Lande2012}, which gauges the spatial extent of the source by taking the difference, $TS_{\rmn{ext}}$ between the TS for a point source signal hypothesis and the extended source signal hypothesis. \citet{Lande2012} found that sources with $TS_{\rmn{ext}}>16$ are confidently ascertained to be spatially extended beyond the LAT point-spread function (PSF).\footnote{In \citet{Lande2012} the authors used a disk to perform their studies.} For \object{Abell~1367} and \object{Abell~3112} we find $TS_{\rmn{ext}}=2.6$ and $TS_{\rmn{ext}}=0.9$, respectively. {{Assuming that the excesses towards all three clusters constitutes point-like emission, we first obtained a better estimate for the location of the excess using the {\tt{gtfindsrc}} tool and then repeated the calculation of TS maps using a finer binning of $0.02\deg$. We report the best-fit values of these excesses along with their uncertainties in Table~\ref{tab:gtfindsrc}.}}
\begin{deluxetable}{lrrr}
\tablewidth{0.9\columnwidth}
\tablecaption{Best-fit positions of excess emission}
\tablehead{\colhead{Host Cluster} & \colhead{R.A.} & \colhead{Dec.} & \colhead{$r_{68}$} \\ 
\colhead{} & \colhead{(\deg)} & \colhead{(\deg)} & \colhead{(\deg)} } 

%% All data must appear between the \startdata and \enddata commands
\startdata
% gtfindsrc
%% A0400 & 44.7 & 5.88 & 0.04 \\
%% A1367 & 176.27 & 19.56 & 0.05 \\
%% A3112 & 49.56 & -44.22 & 0.06 \\
% refined gttsmap
Abell 0400 & $44.68$ & $5.86$ & $0.03$\\
Abell 1367 & $176.25$ & $19.54$ & $0.04$\\
Abell 3112 & $49.56$ & $-44.22$ & $0.02$
\enddata
\tablecomments{\label{tab:gtfindsrc}{We report the best-fit positions from our refined search using the {\tt{gttsmap}} tool that employs a maximum likelihood analysis in order to localize a new point source. The columns from left to right: name of the host cluster, R.A. (J2000), Dec. (J2000) and uncertainty ($r_{68}$). All values are given in degrees.}}
\end{deluxetable}

We also investigate the spectral behavior of the excesses towards Abell~1367 and Abell~3112, by replacing the hadronic CR spectrum with a featureless power-law, such that the flux becomes:
\begin{eqnarray}
\nonumber F_{\gamma}^{\rmn{PL}}\left(E\right)&=& A_\gamma\times \Fexp(>500\mev)\\
&\times& \frac{(1-\Gamma)\times E^{-\Gamma}}{E_{\rmn{max}}^{1-\Gamma}-E_{\rmn{min}}^{1-\Gamma}},
\label{eq:powerlaw}
\end{eqnarray}
where $\Gamma$ is the spectral index, and $E_{\rmn{min}}=500\mev$ and $E_{\rmn{max}}=200\gev$ correspond to the energy range of the analysis. $\Fexp(E>500\mev)$ denotes the expected integral flux above $E_{\rmn{min}}$. $A_\gamma$ corresponds to the scale factor introduced in Section~\ref{sec:scalefactor}. For this test we leave both $A_\gamma$ and $\Gamma$ free to vary. We report the best-fit values of these parameters in Table~\ref{tab:HighTsClusters}. 

\begin{deluxetable*}{lccccc}
%\begin{deluxetable}{lccccc}

\tablewidth{0pt}
\tablecaption{Spectral Model Comparison}
%% correct number of &s.
\tablehead{\colhead{Cluster} & \colhead{$A_\gamma$} & \colhead{TS} & \colhead{$A_\gamma$} & \colhead{$\Gamma$} & \colhead{$TS_{PL}$} \\ 
\colhead{} & \colhead{(hadronic model)} & \colhead{(hadronic model)} & \colhead{(power law)} & \colhead{(power law)} & \colhead{(power law)} } 

%% All data must appear between the \startdata and \enddata commands
\startdata
Abell 1367 & $3^{+2}_{-1}$ & 13.8 & $1.7\pm0.9$ & $1.7\pm0.3$ & 17.3 \\
Abell 3112 & $3\pm2$ & 13.8 & $2.0\pm1.5$ & $1.7\pm0.4$ & 16.1 \\
Abell 400 & $39^{+11}_{-10}$ & 52.7 & $43\pm8$ & $2.3\pm0.2$ & 52.8 \\
\enddata
\tablecomments{Spectral model comparison of clusters that exhibit excess emission. Shown are the best-fit values for $A_\gamma$ along with their associated $TS$ values for the hadronic model and the corresponding values for $A_\gamma$ when replacing the spectrum by a featureless power-law of index $\Gamma$, given by Eq.~\eqref{eq:powerlaw}. The last column indicates the obtained $TS$ value with the power-law fit.}
\label{tab:HighTsClusters}
\end{deluxetable*}
%\end{deluxetable}

For both \object{Abell~1367} and \object{Abell~3112} we find that harder spectral indices are preferred over softer, with a best-fit value for $\Gamma=1.7\pm0.3$ and $\Gamma=1.7\pm0.4$, for Abell~1367 and Abell~3112 respectively.  

While none of the aforementioned tests decisively exclude the attribution of the observed excesses to CR-induced \gr\ emission from the ICM, we note that both \object{Abell~3112} and \object{Abell~1367} are hosts to head-tail radio sources which may be the source of the observed \gr\ emission (see, e.g. \citealt[][]{1982IAUS...97...89G,1987A&A...186L...1G} for \object{Abell~1367} and \citet{Costa1998} for \object{Abell~3112}). {A discussion of supporting multifrequency arguments is given in Appendix~\ref{sec:mwdiscussion}.} 

The excess found in \object{Abell~400} yields a $TS$ value of 52.7 which nominally corresponds to a significance of 7.3\,\ssigma pre-trial (6.7\,\ssigma post-trial). We performed the same tests as in the case of \object{Abell~1367} and \object{Abell~3112} and find that the excess is contained within the cluster virial radius, although the $TS$ map indicates that the emission is about 0.3\deg offset from the cluster center. The difference between the $TS$ evaluated at the cluster center and at the position of the excess is 38. The test for extendedness (centered at the cluster center) yields $TS_{\rmn{ext}}=15.0$, indicating a preference for an extended source, which is likely explained by the offset of the excess. This casts further doubt on the explanation of the excess being ICM emission that moreover would be concentrated towards the cluster center. Fitting the excess with a power-law spectral model as in the case of Abell~3112 and Abell~1367 yields a best-fit value for $\Gamma=2.3\pm0.2$ and $A_\gamma=43\pm8$ with an associated $TS$ value of 52.8, which is similar to that obtained for the hadronic model.
 
In addition to these tests, we searched for source variability using aperture photometry and found no indications of variability on time scales of one month
We note however, that the obtained scale factor for Abell~400, $A_\gamma=39_{-10}^{+11}$, is in strong tension with baseline model expectations and the scale factor constraints derived from other clusters in our sample. If the excess towards Abell~400 constitutes a signal, the calculated upper limit from Abell~400 corresponds to the usual statement that scale factors larger than the upper limit are inconsistent with the data on the stated confidence level. If the excess instead stems from unmodeled background the upper limit means that scale factors larger than the upper limit would make the model more inconsistent with the data than the background-only hypothesis allows at the stated confidence level. In both cases, the result is a conservative upper limit and in addition (as mentioned in Section~\ref{sec:scalefactor}) Abell~400 does not affect the combined upper limit sizeably. Removing Abell~400 from the sample yields a marginally stronger upper limit on the common scale factor for the combined sample of the remaining 49 clusters of $A_\gamma^{UL}=0.40$.

In the conservative CR model of \citet{Pinzke2010}, $A_\gamma = 1$ corresponds to a pion decay \gr\ flux owing to CR protons accelerated at structure formation shocks with a maximal acceleration efficiency and neglecting active CR transport.\footnote{The acceleration efficiency that was used in the simulations was based on observations of supernova remnants and theoretical calculations of diffusive shock acceleration. It is unlikely that there are more efficient mechanisms at work.} Allowing for CR streaming transport would cause a net CR flux to the dilute outer cluster regions and reduce the \gr\ yield for that acceleration efficiency. Additionally, \citet{Pinzke2010} presented an optimistic CR model including effects which enhance the predicted \gr\ yield by a factor of 2-3 depending on the cluster mass.\footnote{A part of the enhancement was due to the numerical limitation of smoothed particle hydrodynamics to excite Kelvin-Helmholtz instabilities and fully mix ram-pressure stripped interstellar medium into the ICM.} However, that model does not account for CRs that are injected by AGNs over the cluster lifetime, which could also produce pions in inelastic collision with the ICM. The total energy dissipation by gravitational shocks exceeds that of AGN for large clusters, making it unlikely that AGN-injected CRs dominate the diffuse \gr\ signal. However, this argument does not apply for smaller CCs (in particular in their central cluster regions) where the AGN appears to dominate the energy budget and could possibly also give rise to observable \gr\ emission \citep[see, e.g.,][ for the interpretation of the low-state of M87 in terms of diffuse pion decay emission while the high state is attributed to jet-induced emission]{2013arXiv1303.5443P}. Hence, the most likely explanation for the possible signal with $A_\gamma \simeq 39$ in Abell~400 is jet-related emission from point sources projected onto or within the cluster. A potential source for the emission in \object{Abell~400} may be the quadruple head-tail system \object{3C~75} which also shows X-ray core emission in the galaxies \citep{1985ApJ...294L..85O,2006A&A...453..433H}. However, given that 3C~75 is located towards the center of \object{Abell~400} and the excess is 0.3\deg offset, which is about eight times larger than the 68\% error radius for a pointsource, this possibility is unlikely. 

\subsection{Individual Upper Limits on the \gr\ Flux}
\label{sec:upperLimits}
Assuming that each cluster in our sample can be modeled according to our description in Section~\ref{sec:cosmic-rays}, one can also derive \emph{individual} limits on $A_{\gamma,i}$ where $i$ refers to the individual cluster. These individual limits can be propagated into flux limits. 

In addition, in order to assess the impact of modeling the clusters in our sample as extended sources, we have derived individual flux upper limits modeling the clusters as point sources.

\begin{figure}
%\plottwo{fluxUL.eps}{fluxUL.eps}
\plotone{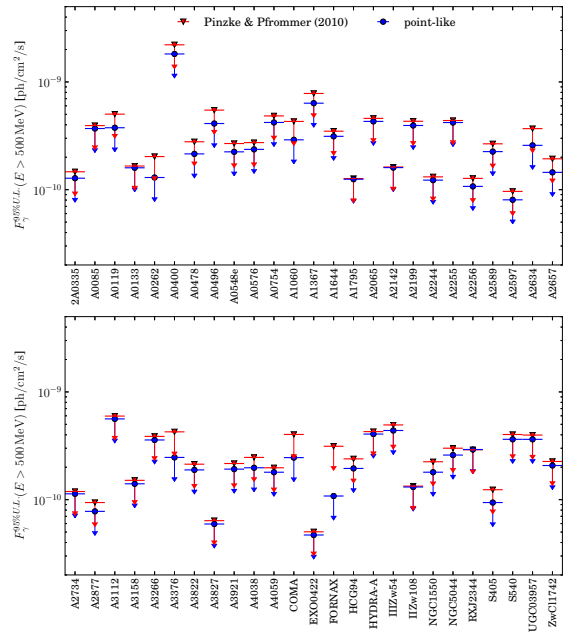}
\caption{(color online)\label{fig:fluxlimits} Shown are the 95\% upper limits on hadronic CR-induced \gr\ flux for each of our 50 galaxy clusters in this analysis. We show the \emph{individually} derived upper limits for both the extended emission (red, downward triangle) and assuming the cluster emission to be point-like (blue, circle).}
\end{figure}

In Figure~\ref{fig:fluxlimits} we show these two cases, contrasting the individually derived \gr flux limits for extended emission from those derived when assuming the cluster emission to be point-like. We tabulate the former for various energy bands along with their associated (pre-trial) $TS$ values in Table~\ref{tab:cluster_limits_individual}. We note that the limit on $A_{\gamma}$, derived from the Coma cluster alone is comparable to the jointly derived limit from the full sample of all 50 clusters in this study, emphasizing its weight in the joint analysis. We investigated the potential dependence of the upper limits on the Galactic diffuse emission model.  In the great majority of cases the limits vary by less than 30\% for the range of models that we considered.  Those clusters for which the dependence on the model was more sensitive are marked in Table \ref{tab:cluster_limits_individual}. However, this sensitivity does not affect the combined limit.

\begin{figure}
%\plottwo{fluxUL.eps}{fluxUL.eps}
\plotone{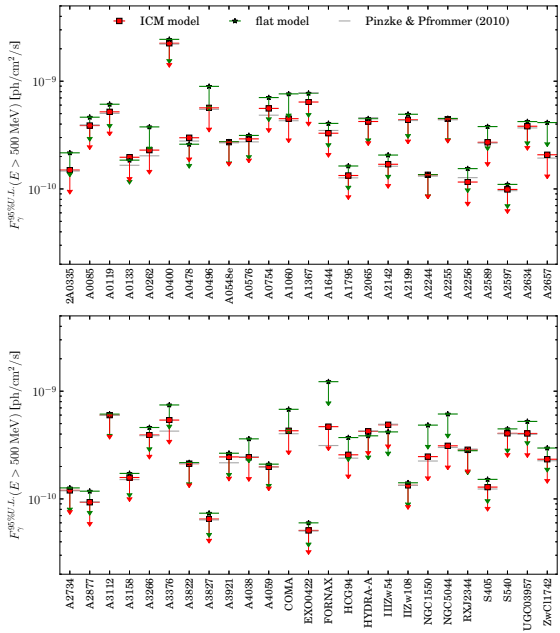}
\caption{(color online)\label{fig:fluxlimits_alt} Same as Figure~\ref{fig:fluxlimits} but for spatial CR profiles following a constant $X_\CR$ profile (ICM model, red squares) and a constant $P_\CR$ profile (flat model, green stars). To allow for an easier comparison we show the limits from the baseline analysis (Figure~\ref{fig:fluxlimits}) in horizontal grey lines.}
\end{figure}

{In Figure~\ref{fig:fluxlimits_alt} we show individual \gr flux upper limits that are derived for the CR profile following: (1) a constant $X_\CR$ profile (ICM model) and (2) a constant $P_\CR$ profile (flat model). We find that the limits derived from the first model are very similar to those with our simulation-based model. On the other hand, the flat model (i.e., constant CR pressure profile) yields less stringent upper limits. We also note that the associated $TS$ values for the flat CR profile are generally marginally higher than either those derived from the thermal ICM or our simulation-based approach. This is expected since the choice of a flat CR profile yields a substantially flatter \gr surface brightness profile, which in turn provides a better fit to the data compared to a cored profile, in particular when considering that the excess emission we report in the previous section is offset from the cluster center. We also note that from the three excesses found, only those towards Abell~3112 and Abell~400 remain with $TS>9$ when using the flat CR profile instead.}

The flux limits we derive substantially improve over previous limits \citep{2010ApJ...717L..71A} due to the increase in data volume and improved modeling of the \gr\ sky as well as improved instrument understanding reflected by the use of reprocessed LAT data with on-orbit calibrations. Finally, we note that in particular for spatially extended clusters such as Coma or Fornax, the limits are substantially weakened with respect to when modeling them as point-like objects.

Motivated by the \Fermi-LAT detection of few bright cluster galaxies {(e.g. 2FGL J0627.1$-$3528 in A3392 and 2FGL J1958.4-3012 in RXC J1958$-$3011)}, a recent stacking study by \citet{Dutson2013} investigated a large sample of galaxy clusters that was selected according to the radio flux of bright cluster galaxies. Although based on a different scientific prior and methodology than our cluster analysis, the determined flux limits can be compared to the point-source upper limits reported here. About a dozen clusters are in common between both studies. However, \citet{Dutson2013} used the respective coordinates of the bright cluster galaxy, which are not necessarily consistent with the cluster center coordinates considered in our studies. Given the LAT PSF \citep{ReprocessingPaper} and the considered ROIs this does not constitute a severe handicap for comparison. The use of different exposures (45 months in Dutson et al. 2013 and 48 months in this work), different source models and, perhaps most particularly noteworthy,  the use of reprocessed LAT data with associated different Galactic and isotropic diffuse models ({{\tt{gal\_2yearp7v6\_v0}}\, vs. \gal\ and {\tt{iso\_p7v6source}}\, vs. \egal\, respectively), as well as different analysis energy thresholds render a strict comparison more problematic. For the majority of common clusters, the limits in Dutson et al. 2013 are marginally less sensitive, as expected regarding the slightly less exposure and the rather moderate changes in the diffuse background models.

However, there are two noticeable exceptions: Abell~85 and Abell~2634 appear to have more constraining upper limits in \citet{Dutson2013}, besides less exposure and a lower analysis threshold. The discrepancies could be explained by differences in the construction of the ROI (treatment of variable sources, sources to be too faint to be in the 2FGL catalog, size of ROI). Taking the respective flux limits at face value, the differences do not amount to more than 35\% between both studies. Our limits on extended cluster emission cannot be meaningfully compared to the point-source upper limits in \citet{Dutson2013} as they constitute alternative scientific priors for a different scientific problem. However, our individual limits on the \gr\ flux, while being specifically derived within the framework of the universal CR model by \citet{Pinzke2010}, can in principle be used to constrain other classes of models.

%\input{tab4.tex}
% baseline vs. pointlike
\begin{deluxetable*}{lrrrrrrrrrr}
%\begin{deluxetable}{lrrrrrrrrrr}

%% Keep a portrait orientation

%% Over-ride the default font size
%% Use Default (12pt)
%\tabletypesize{\tiny}
%\tablenum{4}
%\rotate
%% Use \tablewidth{?pt} to over-ride the default table width.
%% If you are unhappy with the default look at the end of the
%% *.log file to see what the default was set at before adjusting
%% this value.

%% This is the title of the table.
\tablecaption{Individual Flux Upper Limits}
\tablewidth{0pt}
%% This command over-rides LaTeX's natural table count
%% and replaces it with this number.  LaTeX will increment 
%% all other tables after this table based on this number

%% The \tablehead gives provides the column headers.  It
%% is currently set up so that the column labels are on the
%% top line and the units surrounded by ()s are in the 
%% bottom line.  You may add more header information by writing
%% another line between these lines. For each column that requries
%% extra information be sure to include a \colhead{text} command
%% and remember to end any extra lines with \\ and include the 
%% correct number of &s.
\tablehead{
\colhead{Cluster} & \colhead{$A_{\gamma,i}^{UL}$} & \colhead{$F^{UL}_{\gamma,500\mev}$} & \colhead{$F^{UL}_{\gamma,1\gev}$} & \colhead{$F^{UL}_{\gamma,10\gev}$} & \colhead{TS} & \colhead{$A_{\gamma,i}^{UL}$} & \colhead{$F^{UL}_{\gamma,500\mev}$} & \colhead{$F^{UL}_{\gamma,1\gev}$} & \colhead{$F^{UL}_{\gamma,10\gev}$} & \colhead{TS} \\ 
\colhead{} & \colhead{} & \colhead{($\times10^{-10}$)} & \colhead{($\times10^{-11}$)} & \colhead{($\times10^{-12}$)} & \colhead{} & \colhead{} & \colhead{($\times10^{-10}$)} & \colhead{($\times10^{-11}$)} & \colhead{($\times10^{-12}$)} & \colhead{}  \\
\colhead{} & \colhead{extended} & \colhead{extended} & \colhead{extended}  & \colhead{extended} & \colhead{extended} & \colhead{point-like} & \colhead{point-like} & \colhead{point-like}  & \colhead{point-like} & \colhead{point-like}}

%% All data must appear between the \startdata and \enddata commands
\startdata
2A0335 & 0.52 & 1.5 & 6.7 & 3.6 & 0.0 & 0.45 & 1.3 & 5.9 & 3.1 & 0.0 \\
A0085 & 1.35 & 3.9 & 18.0 & 9.5 & 1.0 & 1.26 & 3.7 & 16.9 & 9.0 & 1.3 \\
A0119 & 3.54 & 5.0 & 23.0 & 12.2 & 2.7 & 2.65 & 3.8 & 17.2 & 9.1 & 0.8 \\
A0133 & 2.32 & 1.7 & 7.6 & 4.0 & 0.0 & 2.23 & 1.6 & 7.3 & 3.9 & 0.0 \\
A0262 & 1.67 & 2.0 & 9.3 & 4.9 & 0.0 & 1.07 & 1.3 & 5.9 & 3.1 & 0.0 \\
A0400 & 50.11 & 22.2 & 101.7 & 53.6 & 52.7 & 41.20 & 18.3 & 83.6 & 44.1 & 37.7 \\
A0478\tablenotemark{a} & 1.14 & 2.8 & 12.7 & 6.7 & 0.0 & 0.88 & 2.1 & 9.8 & 5.2 & 0.0 \\
A0496\tablenotemark{a} & 1.94 & 5.5 & 25.2 & 13.3 & 1.9 & 1.46 & 4.1 & 18.9 & 10.0 & 0.5 \\
A0548e & 10.78 & 2.7 & 12.3 & 6.5 & 0.1 & 9.02 & 2.2 & 10.3 & 5.4 & 0.0 \\
A0576 & 4.39 & 2.7 & 12.6 & 6.6 & 0.2 & 3.80 & 2.4 & 10.9 & 5.7 & 0.3 \\
A0754 & 2.86 & 4.8 & 22.2 & 11.7 & 1.1 & 2.49 & 4.2 & 19.3 & 10.2 & 0.7 \\
A1060 & 1.89 & 4.3 & 19.7 & 10.4 & 0.4 & 1.28 & 2.9 & 13.3 & 7.0 & 0.1 \\
A1367 & 4.08 & 7.8 & 35.9 & 19.0 & 13.8 & 3.31 & 6.4 & 29.1 & 15.4 & 11.2 \\
A1644 & 2.69 & 3.5 & 16.0 & 8.5 & 0.2 & 2.41 & 3.1 & 14.3 & 7.6 & 0.4 \\
A1795\tablenotemark{a} & 0.42 & 1.3 & 5.8 & 3.1 & 0.0 & 0.42 & 1.3 & 5.7 & 3.0 & 0.0 \\
A2065 & 5.00 & 4.6 & 21.1 & 11.2 & 2.3 & 4.69 & 4.3 & 19.8 & 10.5 & 2.2 \\
A2142 & 0.47 & 1.6 & 7.4 & 3.9 & 0.0 & 0.46 & 1.6 & 7.3 & 3.9 & 0.0 \\
A2199 & 1.45 & 4.3 & 19.8 & 10.5 & 1.9 & 1.32 & 4.0 & 18.1 & 9.6 & 1.7 \\
A2244 & 1.74 & 1.3 & 6.0 & 3.2 & 0.0 & 1.63 & 1.2 & 5.6 & 3.0 & 0.0 \\
A2255 & 5.21 & 4.4 & 20.2 & 10.6 & 5.2 & 4.99 & 4.2 & 19.3 & 10.2 & 6.3 \\
A2256 & 0.50 & 1.3 & 5.8 & 3.1 & 0.0 & 0.42 & 1.1 & 4.9 & 2.6 & 0.0 \\
A2589\tablenotemark{a} & 3.96 & 2.7 & 12.2 & 6.5 & 0.0 & 3.36 & 2.3 & 10.4 & 5.5 & 0.0 \\
A2597 & 1.27 & 1.0 & 4.4 & 2.3 & 0.0 & 1.06 & 0.8 & 3.7 & 2.0 & 0.0 \\
A2634 & 5.95 & 3.7 & 17.0 & 8.9 & 0.3 & 4.16 & 2.6 & 11.9 & 6.2 & 0.0 \\
A2657\tablenotemark{a} & 2.34 & 1.9 & 8.9 & 4.7 & 0.0 & 1.75 & 1.4 & 6.6 & 3.5 & 0.0 \\
A2734 & 2.71 & 1.2 & 5.5 & 2.9 & 0.0 & 2.58 & 1.1 & 5.2 & 2.7 & 0.0 \\
A2877 & 1.87 & 0.9 & 4.3 & 2.3 & 0.0 & 1.55 & 0.8 & 3.6 & 1.9 & 0.0 \\
A3112 & 5.37 & 6.0 & 27.3 & 14.4 & 13.8 & 5.06 & 5.6 & 25.7 & 13.6 & 12.8 \\
A3158 & 1.26 & 1.5 & 6.9 & 3.7 & 0.0 & 1.17 & 1.4 & 6.4 & 3.4 & 0.0 \\
A3266 & 1.24 & 3.9 & 17.7 & 9.4 & 2.2 & 1.15 & 3.6 & 16.4 & 8.7 & 3.6 \\
A3376 & 6.20 & 4.3 & 19.5 & 10.3 & 0.0 & 3.59 & 2.5 & 11.3 & 6.0 & 0.0 \\
A3822 & 4.77 & 2.1 & 9.8 & 5.2 & 0.0 & 4.22 & 1.9 & 8.7 & 4.6 & 0.0 \\
A3827 & 0.66 & 0.6 & 2.9 & 1.5 & 0.0 & 0.61 & 0.6 & 2.7 & 1.4 & 0.0 \\
A3921\tablenotemark{a} & 4.63 & 2.2 & 9.9 & 5.2 & 0.1 & 4.11 & 1.9 & 8.8 & 4.6 & 0.0 \\
A4038 & 1.84 & 2.5 & 11.3 & 6.0 & 0.1 & 1.48 & 2.0 & 9.1 & 4.8 & 0.0 \\
A4059 & 2.04 & 2.0 & 9.1 & 4.8 & 0.0 & 1.86 & 1.8 & 8.2 & 4.4 & 0.0 \\
COMA\tablenotemark{a} & 0.35 & 4.0 & 18.5 & 9.8 & 0.7 & 0.22 & 2.5 & 11.3 & 6.0 & 0.1 \\
EXO0422 & 0.70 & 0.5 & 2.3 & 1.2 & 0.0 & 0.65 & 0.5 & 2.2 & 1.1 & 0.0 \\
FORNAX & 3.73 & 3.1 & 14.3 & 7.6 & 0.0 & 1.29 & 1.1 & 4.9 & 2.6 & 0.0 \\
HCG94 & 6.34 & 2.4 & 11.0 & 5.8 & 0.0 & 5.17 & 2.0 & 9.0 & 4.7 & 0.0 \\
HYDRA-A & 2.57 & 4.3 & 19.6 & 10.4 & 3.1 & 2.44 & 4.1 & 18.6 & 9.8 & 3.2 \\
IIIZw54 & 10.97 & 5.0 & 22.7 & 12.0 & 0.5 & 9.74 & 4.4 & 20.2 & 10.6 & 0.3 \\
IIZw108 & 3.31 & 1.3 & 6.1 & 3.2 & 0.0 & 3.25 & 1.3 & 6.0 & 3.2 & 0.0 \\
NGC1550 & 3.37 & 2.3 & 10.3 & 5.5 & 0.0 & 2.70 & 1.8 & 8.3 & 4.4 & 0.0 \\
NGC5044 & 4.96 & 3.0 & 13.8 & 7.3 & 0.0 & 4.29 & 2.6 & 11.9 & 6.3 & 0.0 \\
RXJ2344 & 5.07 & 2.9 & 13.3 & 7.0 & 1.3 & 5.10 & 2.9 & 13.4 & 7.1 & 2.3 \\
S405\tablenotemark{a} & 3.11 & 1.2 & 5.7 & 3.0 & 0.0 & 2.36 & 0.9 & 4.3 & 2.3 & 0.0 \\
S540 & 11.47 & 4.0 & 18.5 & 9.7 & 1.2 & 10.35 & 3.6 & 16.7 & 8.8 & 0.8 \\
UGC03957 & 8.00 & 4.0 & 18.2 & 9.6 & 0.4 & 7.31 & 3.6 & 16.6 & 8.8 & 0.3 \\
ZwCl1742 & 1.94 & 2.3 & 10.4 & 5.5 & 0.0 & 1.77 & 2.1 & 9.5 & 5.0 & 0.0 \\
\enddata

%% Include any \tablenotetext{key}{text}, \tablerefs{ref list},
%% or \tablecomments{text} between the \enddata and 
%% \end{deluxetable*} commands

%% General table comment marker
\tablecomments{The columns contain from left to right the 95\% upper limit on $A_{\gamma,i}$ for each cluster, the derived flux upper limit above (500\mev, 1\gev, and 10\gev), the associated $TS$ value as well as the same quantities assuming the clusters to be modeled by a point source at the cluster position as given in Table~\ref{tab:cluster_overview_all}. Fluxes are given in $\rmn{ph\,s^{-1}\,cm^{-2}}$.}
\tablenotetext{a}{The upper limits on $A_{\gamma,i}$ derived using our alternative diffuse models (Section~\ref{sec:diffuse}) for these clusters varied by more than 30\% relative to those obtained using the standard diffuse emission model.}
%% No \tablerefs indicated
\label{tab:cluster_limits_individual}
\end{deluxetable*}
%\end{deluxetable}

\begin{deluxetable*}{lrrrrrrrrrr}
%\begin{deluxetable}{lrrrrrrrrrr}

%% Keep a portrait orientation

%% Over-ride the default font size
%% Use Default (12pt)
%\tabletypesize{\tiny}
%% \tablenum{5}
%% \rotate
%% Use \tablewidth{?pt} to over-ride the default table width.
%% If you are unhappy with the default look at the end of the
%% *.log file to see what the default was set at before adjusting
%% this value.

%% This is the title of the table.
\tablecaption{Individual Flux Upper Limits (Alternative CR Profiles)}
\tablewidth{0pt}
%% This command over-rides LaTeX's natural table count
%% and replaces it with this number.  LaTeX will increment 
%% all other tables after this table based on this number

%% The \tablehead gives provides the column headers.  It
%% is currently set up so that the column labels are on the
%% top line and the units surrounded by ()s are in the 
%% bottom line.  You may add more header information by writing
%% another line between these lines. For each column that requries
%% extra information be sure to include a \colhead{text} command
%% and remember to end any extra lines with \\ and include the 
%% correct number of &s.
\tablehead{
\colhead{Cluster} & \colhead{$A_{\gamma,i}^{UL}$} & \colhead{$\Fexp$} & \colhead{$F^{UL}_{\gamma,500\mev}$} & \colhead{$TS$} & \colhead{$A_{\gamma,i}^{UL}$} & \colhead{$\Fexp$} & \colhead{$F^{UL}_{\gamma,500\mev}$} & \colhead{$TS$} \\ 
\colhead{Spatial Model} & \colhead{ICM} & \colhead{ICM} & \colhead{ICM} & \colhead{ICM} & \colhead{flat} & \colhead{flat} & \colhead{flat} & \colhead{flat} }

%% All data must appear between the \startdata and \enddata commands
\startdata
2A0335 & 0.85 & 1.8 & 1.5 & 0.0 & 5.14 & 0.4 & 2.2 & 0.0 \\
A0085 & 1.64 & 2.4 & 3.9 & 0.6 & 6.15 & 0.8 & 4.6 & 1.1 \\
A0119 & 3.91 & 1.3 & 5.2 & 2.7 & 9.55 & 0.6 & 6.1 & 3.0 \\
A0133 & 3.83 & 0.5 & 2.0 & 0.0 & 15 & 0.1 & 1.9 & 0.0 \\
A0262 & 2.39 & 1.0 & 2.3 & 0.0 & 5.63 & 0.7 & 3.8 & 0.0 \\
A0400 & 59.23 & 0.4 & 22.5 & 52.9 & 95.1 & 0.3 & 24.5 & 61.4 \\
A0478 & 1.57 & 1.9 & 3.0 & 0.0 & 6.9 & 0.4 & 2.6 & 0.0 \\
A0496 & 2.74 & 2.1 & 5.7 & 2.3 & 14.66 & 0.6 & 9.0 & 5.4 \\
A0548e & 12.8 & 0.2 & 2.7 & 0.1 & 17.91 & 0.2 & 2.7 & 0.1 \\
A0576 & 5.25 & 0.6 & 2.9 & 0.3 & 15.01 & 0.2 & 3.1 & 0.1 \\
A0754 & 3.8 & 1.5 & 5.6 & 2.3 & 19.61 & 0.4 & 7.1 & 2.8 \\
A1060 & 2.63 & 1.7 & 4.5 & 0.4 & 13.74 & 0.6 & 7.7 & 0.1 \\
A1367 & 3.68 & 1.8 & 6.4 & 5.6 & 18.64 & 0.4 & 7.7 & 5.0 \\
A1644 & 2.77 & 1.2 & 3.3 & 0.0 & 7.67 & 0.5 & 4.1 & 0.0 \\
A1795 & 0.58 & 2.3 & 1.3 & 0.0 & 5.03 & 0.3 & 1.6 & 0.0 \\
A2065 & 4.99 & 0.8 & 4.2 & 1.5 & 26.23 & 0.2 & 4.5 & 1.5 \\
A2142 & 0.57 & 3.0 & 1.7 & 0.0 & 2.61 & 0.8 & 2.1 & 0.0 \\
A2199 & 1.84 & 2.4 & 4.4 & 1.9 & 7.09 & 0.7 & 4.9 & 1.2 \\
A2244 & 2.23 & 0.6 & 1.4 & 0.0 & 6.86 & 0.2 & 1.4 & 0.0 \\
A2255 & 5.67 & 0.8 & 4.5 & 5.1 & 11.27 & 0.4 & 4.5 & 4.4 \\
A2256 & 0.49 & 2.4 & 1.2 & 0.0 & 2.13 & 0.7 & 1.6 & 0.0 \\
A2589 & 5.06 & 0.5 & 2.7 & 0.0 & 21.48 & 0.2 & 3.8 & 0.0 \\
A2597 & 1.97 & 0.5 & 1.0 & 0.0 & 12.69 & 0.1 & 1.1 & 0.0 \\
A2634 & 6.84 & 0.6 & 3.8 & 0.4 & 15.34 & 0.3 & 4.2 & 0.3 \\
A2657 & 3.01 & 0.7 & 2.1 & 0.0 & 27.51 & 0.2 & 4.1 & 0.1 \\
A2734 & 3.18 & 0.4 & 1.2 & 0.0 & 7.4 & 0.2 & 1.3 & 0.0 \\
A2877 & 2.14 & 0.4 & 0.9 & 0.0 & 16.41 & 0.1 & 1.2 & 0.0 \\
A3112 & 7.57 & 0.8 & 6.0 & 13.8 & 29.98 & 0.2 & 6.1 & 13.4 \\
A3158 & 1.51 & 1.1 & 1.6 & 0.0 & 3.78 & 0.5 & 1.7 & 0.0 \\
A3266 & 1.34 & 2.9 & 3.9 & 2.1 & 6.17 & 0.7 & 4.6 & 1.7 \\
A3376 & 8.43 & 0.6 & 5.4 & 0.5 & 28.36 & 0.3 & 7.5 & 1.2 \\
A3822 & 5.27 & 0.4 & 2.1 & 0.0 & 9.07 & 0.2 & 2.2 & 0.0 \\
A3827 & 0.72 & 0.9 & 0.6 & 0.0 & 3.91 & 0.2 & 0.7 & 0.0 \\
A3921 & 5.96 & 0.4 & 2.5 & 0.6 & 19.58 & 0.1 & 2.7 & 0.6 \\
A4038 & 2.49 & 1.0 & 2.4 & 0.1 & 11.23 & 0.3 & 3.6 & 0.5 \\
A4059 & 2.66 & 0.7 & 2.0 & 0.1 & 11.15 & 0.2 & 2.1 & 0.0 \\
COMA & 0.41 & 10.4 & 4.3 & 0.9 & 1.37 & 4.9 & 6.8 & 2.2 \\
EXO0422 & 0.92 & 0.6 & 0.5 & 0.0 & 4.83 & 0.1 & 0.6 & 0.0 \\
FORNAX & 7.19 & 0.7 & 4.7 & 0.5 & 74.75 & 0.2 & 12.3 & 1.8 \\
HCG94 & 8.44 & 0.3 & 2.6 & 0.0 & 23.26 & 0.2 & 3.7 & 0.0 \\
HYDRA-A & 3.68 & 1.2 & 4.2 & 3.0 & 18.86 & 0.2 & 3.9 & 1.2 \\
IIIZw54 & 12.94 & 0.4 & 4.9 & 0.4 & 52.7 & 0.1 & 4.2 & 0.0 \\
IIZw108 & 3.66 & 0.4 & 1.3 & 0.0 & 6.43 & 0.2 & 1.4 & 0.0 \\
NGC1550 & 5.38 & 0.5 & 2.5 & 0.0 & 25.85 & 0.2 & 4.9 & 0.0 \\
NGC5044 & 9.95 & 0.3 & 3.1 & 0.0 & 61.79 & 0.1 & 6.1 & 0.1 \\
RXJ2344 & 5.79 & 0.5 & 2.9 & 1.3 & 30.83 & 0.1 & 2.8 & 0.4 \\
S405 & 3.5 & 0.4 & 1.3 & 0.0 & 6.16 & 0.2 & 1.5 & 0.0 \\
S540 & 14.76 & 0.3 & 4.1 & 1.2 & 48.07 & 0.1 & 4.5 & 1.0 \\
UGC03957 & 10.74 & 0.4 & 4.1 & 0.5 & 75.34 & 0.1 & 5.3 & 0.6 \\
ZwCl1742 & 2.34 & 1.0 & 2.3 & 0.0 & 12.66 & 0.2 & 3.0 & 0.0 \\
\enddata

%% Include any \tablenotetext{key}{text}, \tablerefs{ref list},
%% or \tablecomments{text} between the \enddata and 
%% \end{deluxetable*} commands

%% General table comment marker
\tablecomments{Individual flux predictions and upper limits for alternative CR profiles for photon energies above 500\mev. Columns 2-5 refer to the individual scale factor, the flux prediction above 500\mev ($\Fexp$), the derived upper limit on the \gr flux and the individual $TS$ value for the ICM model, while columns 6-9 represent the values obtained for the flat CR model. All photon fluxes are given in $10^{-10}\,\rmn{ph\,s^{-1}\,cm^{-2}}$.}
%% No \tablerefs indicated
\label{tab:cluster_limits_individual_altCR}
\end{deluxetable*}
%\end{deluxetable}

\subsection{CR-to-thermal Pressure Ratio $\langle X_\CR \rangle$}

We show the resulting upper limits on the CR-to-thermal pressure ratio, $\langle X_\CR \rangle$ in Figure \ref{fig:XCRlimits}. These numbers were obtained by scaling the $\langle X_{\CR}\rangle$ values in Table~\ref{tab:cluster_overview_all} with the limit on $A_{\gamma}$ from Section~\ref{sec:scalefactor}. This procedure assumes universality of the CR distribution as suggested by hydrodynamical cosmological simulations of clusters \citep{Pinzke2010} and implicitly asserts that the active CR transport does not appreciably modify the spatial distribution of the CRs. This is justified since the impact of CR streaming on the CR distribution of a cosmological cluster is not clear to date. Depending on the microscopic plasma physics that sets the CR streaming speed (i.e., competing damping mechanisms of the CR Alfv{\'e}n waves) and the macroscopic distribution of cluster magnetic fields \citep[][for evidence of radial bias of the magnetic geometry]{2010NatPh...6..520P, 2011ApJ...740...81R}, CR streaming could either be a perturbation to the peaked advection-dominated CR distribution or cause a substantial flattening by a substantial net outward CR flux \citep{2011A&A...527A..99E,2013arXiv1303.4746W}. 

The median upper limit on the CR pressure ratio is $\langle X_\CR \rangle < 0.006$ for the combined sample and the NCC subsample within $R_{HL}$. Those constraints are relaxed to $\langle X_\CR \rangle < 0.012$ (combined sample) and $\langle X_\CR \rangle < 0.013$ (NCC) within $R_{200}$. For CC clusters, this limit is less stringent, yielding $\langle X_\CR \rangle < 0.008$ and $\langle X_\CR \rangle < 0.014$ within $R_{HL}$ and $R_{200}$, respectively. 

\begin{figure}[tbp]
%\epsscale{.95}
\plotone{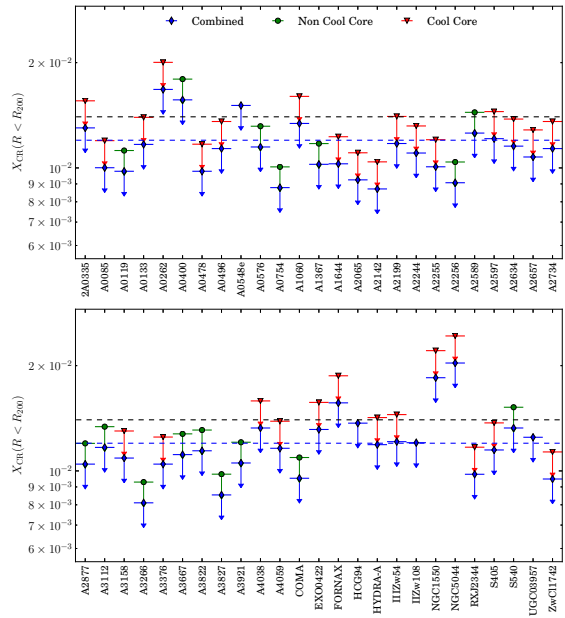}
%\plottwo{XCR_UL.eps}{XCR_UL.eps}
\caption{(color online)\label{fig:XCRlimits} Individual 95\% upper limits on $X_\CR$ for each of our 50 galaxy clusters in this analysis assuming the \emph{jointly derived} scale factor we obtain in our analysis for the full sample (blue, diamond), CC clusters (red, downward triangle) and NCC clusters (green, circle). The dashed lines represent the median upper limit for the combined (blue) while the median upper limits for CC and NCC are the same (shown in black).}
\end{figure}

These limits are more constraining than previous limits on the CR pressure that were obtained through flux upper limits on individual objects, in particular for individual limits on clusters using the initial 18 months of LAT data \citep{2010ApJ...717L..71A} as well as improving those constraints that use 4 years of \Fermi data on Coma, which yield $\langle X_\CR\rangle< 0.017$ \citep{2012ApJ...757..123A}, provided the CR universality assumption holds (assuming both universality as well as the scaling relation we adopt throughout our work, characterized through $A_\gamma$ for the combined sample, for Coma we find $\langle X_\CR\rangle< 0.011$). The most suitable cluster target for CR-induced \gr\ emission, the Perseus cluster, cannot be used to competitively constrain the CR pressure using \Fermi-LAT data since the \Fermi-LAT and MAGIC collaborations detected the central radio galaxy NGC~1275 in \grs\ in the energy range from 300\mev\ to $>300\gev$ (\Fermi: 300\mev-300\gev, MAGIC: $>200\gev$) \citep{2009ApJ...699...31A,2012A&A...539L...2A}. Non-observations of \grs\ from the Perseus cluster {\em above} these energies by the MAGIC Collaboration \citep{2012A&A...541A..99A,2010ApJ...710..634A} provide limits similar to those obtained from analyzing LAT observations of Coma, $\langle X_\CR\rangle< 0.017$ alone \citep{2012ApJ...757..123A,2013arXiv1312.1493Z}.
 
Our limits on $X_\mathrm{CR}$ probe the entire ICM and are much more constraining in comparison to limits on the non-thermal pressure contribution of the central ICM in several nearby cD galaxies that have been derived by comparing the gravitational potentials inferred from stellar and globular cluster kinematics and from assuming hydrostatic equilibrium of the X-ray emitting gas \citep{2008MNRAS.388.1062C,2010MNRAS.404.1165C}. Depending on the adopted estimator of the optical velocity dispersion, they find a non-thermal pressure bias of $X_\mathrm{nt}=P_\mathrm{nt}/P_\mathrm{th}\approx0.21-0.29$, which probes the cumulative non-thermal pressure contributed by CRs, magnetic fields, and unvirialized motions. It is, however, conceivable that those central regions of CC clusters, which probe the enrichment of non-thermal components as a result of AGN feedback \citep[e.g.,][]{2013arXiv1303.5443P}, are characterized by a larger CR pressure contribution in comparison to the bulk of the ICM that probes CRs accelerated by shocks associated with the growth of structure and magnetic fields that are reprocessed by these shocks.

{Complementary limits on CRs are also derived from radio (synchrotron) observations \citep{2004A&A...413...17P,Brunetti2007}. Assuming a central cluster magnetic field of $B\sim1\mu G$ and CR spectral indices $\alpha=2.1$--2.4, this approach allows the CR energy to be constrained to a few percent while the limits are less stringent for steeper spectra and lower magnetic fields \citep{2009A&A...502..437A}.}

\subsection{Hadronic Injection Efficiency}
\label{sec:inj_eff}

{ The distributions of CRs within the virial regions of clusters are built up
  from shocks during the cluster assembly
  \citep{2003ApJ...593..599R,2006MNRAS.367..113P,2013MNRAS.435.1061P}. While
  strong shocks are responsible for the high-energy population of CRs that potentially could be visible at TeV $\gamma$-ray energies, intermediate Mach-number shocks with
  $\mathcal{M}\simeq3$--$4$ build up the CR population at GeV energies
  \citep{Pinzke2010}, which can be constrained through \Fermi observations. The
  normalization of the CRs and \gr\ flux scale with the acceleration efficiency
  at the shocks of corresponding strength. Our fiducial CR model
  \citep{Pinzke2010} uses a simplified model for CR acceleration
  \citep{2007A&A...473...41E} in which the efficiency rises steeply with Mach
  number for weak shocks and saturates already at shock Mach numbers
  $\mathcal{M}\gtrsim3$.  Observations of supernova remnants
  \citep{2009Sci...325..719H} and theoretical studies
  \citep{2005APh....24...75J} suggest a value for the saturated acceleration
  efficiency of $\zeta_{\rmn{p,max}}\simeq0.5$, i.e., the fraction of
  shock-dissipated energy that is deposited in CR ions. Following these
  optimistic predictions, the model of \citet{Pinzke2010} assumes this value for
  the saturated efficiency, which serves as input to our analysis.  This model
  provides a plausible upper limit for the CR contribution from structure
  formation shocks in galaxy clusters since more elaborate models of CR
  acceleration predict even lower efficiencies for the Mach-number range
  $\mathcal{M}\simeq3$--$4$ of relevance here \citep[][see also
  Appendix~\ref{sec:acc_eff} for a more detailed
  discussion]{2013ApJ...764...95K}.  If we scale the CR pressure contribution
  linearly with the maximum acceleration efficiency, using the previously
  derived limit on $A_\gamma$, we find $\zeta_{\rmn{p,max}}^{UL}=21\%$ for the
  combined sample while the CC- and NCC-subsamples yield 25\% and 24\% maximum
  acceleration efficiency, respectively. We note that this constrains the
  maximum acceleration efficiency {\em only} in the simplified model adopted
  here. For different acceleration models, these upper limits provide
  conservative constraints on the acceleration efficiency at intermediate
  strength shocks of Mach number $\mathcal{M}\simeq3$--$4$, a regime
  complementary to that studied at supernova remnant shocks.}

However, these conclusions rely on two major assumptions, namely CR universality and the absence of efficient CR transport relative to the plasma rest frame.  The latter assumption hypothesizes that CRs are tied to the gas via small-scale tangled magnetic fields, which implies that they are only advectively transported and that we can neglect the CR streaming and diffusive transport relative to the rest frame of the gas. Early work on this topic suggests that such CR transport processes are at work in clusters and cause a flattening of radial CR profiles{{ that can significantly reduce the radio and \gr emission at high energies probed by Cherenkov telescopes but remains largely unaffected at lower energies probed by }\Fermi} \citep{2013arXiv1303.4746W}. 
Moreover, different formation histories of clusters cause spatial variations of the CR distribution and hence a deviation from a universal distribution \citep{Pinzke2010}. To date, there is no consensus about the size of these effects and more work is needed to fully quantify them.

\section{Systematic Uncertainties}
\label{sec:systematics}
\subsection{Choice of Fiducials: Binning, Region Size, Free Sources}
\subsubsection{Binning}
While the number of spectral bins is the same for the whole sample (nominally 18 bins), the number of spatial bins varies with the ROI size, because the bin width is constant (0.1\deg). In addition, the extended templates used to model the cluster emission are also binned. To check the effect of binning, we vary the nominal values by 50\%. Aside from potentially increasing the computation time for larger numbers of spatial bins we find that our choice of binning does not change the results by more than $\sim1\%$. Similarly we find that varying the number of spectral bins changes the resulting limits on $A_\gamma$ by at most 5\%.

\subsubsection{Region Size}
We use ROIs of varying sizes, ranging from 8--16\deg in radius (Table~\ref{tab:roidefinition}). To make sure that this choice does not introduce any significant bias, we compare the fitted values for all free parameters for the null hypothesis in these smaller regions with ROIs with 25\% larger radii and find variations of these values which are less than 3\% with respect to larger regions. However, we note that larger regions allow a more stringent determination of the background model which is reflected by smaller uncertainties on the Galactic and isotropic diffuse components than for the case of smaller regions (compare to error bars in Figure~\ref{fig:diffuse_normalizations}). 

\subsubsection{Free Sources}
We choose to use the 2-year source list to model data collected during 4 years of LAT observations.  While we ensure that residual excesses are mitigated by allowing the normalizations of the known point sources to vary, the choice of leaving the normalizations of sources within 4\deg of each cluster to vary freely is somewhat arbitrary. Freeing the normalizations of only those sources within $\theta_{200}+1\deg$ does not change our results on $A_{\gamma}$ by more than 10\%. %Regardless of which of these two methods are used, we ensure that there are no unmodeled point sources in the 4 year dataset that are not in the 2 year dataset.%We find that this choice does not change our results by more than 10\%. We consider our choice as conservative and robust as it allows to attribute excess emission to already observed point sources and by using {\tt MINOS} we properly take into account correlations between the involved free parameters.
\subsection{Event Classes and Instrument Response Functions}
The IRFs consist of three separate parts \citep[see][for details]{Instrumentpaper}: the effective area which has an associated uncertainty of at most $\sim10\%$ in the energy range we consider, the PSF whose uncertainty can be conservatively estimated to be $\sim15\%$, and the energy dispersion (whose uncertainty is negligible for this analysis). Using bracketing IRFs (see Sections 5.7.1 and 6.5.1 in \citet{Instrumentpaper}) to quantify these uncertainties we find that while individual ROIs may show variations of up to $\sim21\%$, the effect on the combined scale factors and quantities derived from it is less than 7\%.

\subsection{Diffuse Emission}
\label{sec:diffuse}
Spatial residuals due to mismodeling of the large-scale Galactic diffuse foreground emission may be misinterpreted in terms of an extended \gr\ excess. We compare results derived using the standard diffuse emission model adopted for the baseline analysis (based on empirical fits of multiple spatial templates to \gr\ data) to results obtained when using a set of eight alternative diffuse emission models that were created using a different methodology with respect to the standard diffuse emission model.\footnote{We also use a different set of isotropic diffuse templates that were created in conjunction with the alternative Galactic diffuse templates.}

We chose these models to represent the most important parameters scanned in \citet{DiffusePaper2}, in particular, CR source distribution, halo size and spin temperature. We summarize the properties of the alternative models in Table 7, and refer readers to \citet{2013arXiv1304.1395D} for details. The models we employed were tuned to the {\tt{P7REP}} data. Although the models were created such that different components along the line of sight could be fit separately, we only adopted a free overall normalization since at high Galactic latitudes the vast majority of the gas resides in the neighborhood of the solar system. Moreover, having different components as additional degrees of freedom in the fit makes a comparison of the $TS$ values with the baseline analysis more difficult.

\begin{deluxetable}{cccc}
\tablewidth{\columnwidth}
%\tabletypesize{\scriptsize}
\tablecaption{Alternative Models for Galactic Diffuse Emission}
\tablehead{\colhead{Label} & \colhead{CR Source Distribution} & \colhead{Halo Size} & \colhead{Spin Temperature} \\ 
\colhead{} & \colhead{} & \colhead{(kpc)} & \colhead{(K)} } 

%% All data must appear between the \startdata and \enddata commands
\startdata
A & Lorimer & 10 & $10^5$ \\
B & Lorimer & 10 & 150 \\
C & Lorimer & 4 & $10^5$ \\
D & Lorimer & 4 & 150 \\
E & SNR & 10 & $10^5$ \\
F & SNR & 10 & 150 \\
G & SNR & 4 & $10^5$ \\
H & SNR & 4 & 150 \\
\enddata

%% Include any \tablenotetext{key}{text}, \tablerefs{ref list},
%% or \tablecomments{text} between the \enddata and 
%% \end{deluxetable*} commands

%% General table comment marker
\tablecomments{\label{tab:GALPROP}Overview of the alternative diffuse models used for assessing the systematic uncertainties in the model for the Galactic diffuse emission. We chose to vary the 3 most important input parameters that were found in scanning the parameter space in \citet{DiffusePaper2}. }
\end{deluxetable}

%The GALPROP H~{\sc i}, H~{\sc ii}, and CO output maps were binned into four Galactocentric annuli from 0--4~kpc, 4--8~kpc, 8--10~kpc and 10--30~kpc. To minimize bias in the a-priori assumptions on the CR injection spectra and the proton CR source distribution the models were adjusted to the gamma-ray data in a likelihood fit to 2 years of \Fermi-LAT data. The spectra of each annulus was multiplied with a log-parabola function. To allow for possible CR spectral variations within the annuli the CO annuli had a separate log-parabola function from the H~{\sc i} and H~{\sc ii} annuli. The inverse Compton (IC) template from GALPROP was also adjusted with a log-parabola function to account for spectral variations in the electron distribution. Since the IC template is not split into Galactocentric bins we could not fit the CR electron source distribution. 

%In addition to the interstellar emission model, we also included an isotropic template and templates for Loop I \citep{CasandjianLoopI} and the \Fermi Lobes \citep{SuFinkbeiner2010}. The template for Loop I is based on the geometrical model of \citet{Wolleben2007} while the lobes are assumed to be uniform with edges defined by $R = R_0 \abs{\cos\theta}$.

%Here, $\theta$ is defined as the angle between the positive $z$ axis and the galactic plane (spherical coordinates) and $R_0$ was chosen to be 8~kpc to roughly match the shape of the lobes. When adjusting the fit, point sources from the 2FGL \citep{Abdo2010} were included in the fit.

We emphasize that these 8 models do not span the complete uncertainty of the systematics involved with interstellar emission modeling. They do not even encompass the full uncertainty in the input parameters that are varied. The resulting uncertainty should therefore only be considered as one indicator of the systematic uncertainty due to interstellar emission modeling. 
% maybe we can use a shorter paragraph...
The tests we performed using these additional models considered a different energy range, from {500\mev--100\gev}, because the alternative models were not derived for higher energies. However, since events at low energies dominate the fit, this difference is negligible. We have explicitly verified this by repeating the baseline analysis up to 100\gev\, and found that the differences between the computed combined scale factors are $<1\%$. 

In Figure~\ref{fig:diffuse_comparison} we show how the combined upper limit on the scale factor, $A_{\gamma}^{UL}$ varies for the alternative models with respect to our standard model as well as how the significance of the excesses observed in Abell~1367, Abell~3112 and Abell~400 changes when using the alternative models. 

\begin{figure}[tbp]
%\epsscale{.95}
\plotone{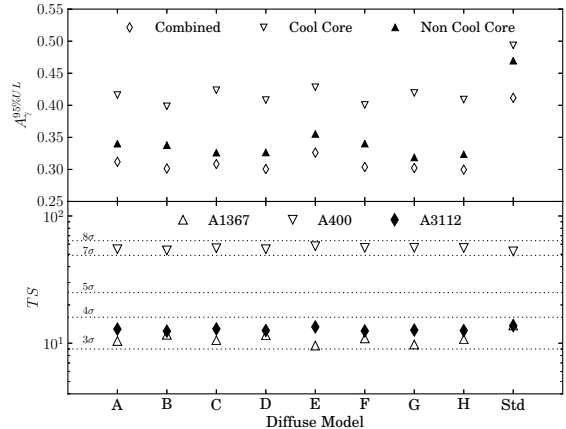}
\caption{\label{fig:diffuse_comparison} In the upper panel we show the 95\% combined upper limit on $A_\gamma$ for the respective (sub)samples for the alternative diffuse models (A-H, see Table~\ref{tab:GALPROP} and the accompanying text for details). In the bottom panel we show the associated $TS$ values for the three clusters that exhibit significant excess emission.}
\end{figure}

The spread in limits for the scale factor is rather small between the alternative diffuse models, but the choice of using the standard diffuse model versus any of the alternative diffuse models can affect the inferred limits for the different cluster samples by $20-30\%$.
Comparing the $TS$ values for the three clusters indicates small variations across the alternative models for Abell~400 and Abell~3112 and Abell~1367. We have repeated this procedure for the derivation of the individual flux limits (see Section~\ref{sec:upperLimits}) and find that the majority of our clusters follow the same trend. We have marked the clusters which show variations beyond $30\%$ in Table~\ref{tab:cluster_limits_individual}.

We summarize the systematic uncertainties discussed in this section in Table~\ref{tab:uncertainties} and note that the main source of uncertainty is the accurate modeling of the foreground Galactic diffuse emission. 

\begin{deluxetable*}{lcc}
%\begin{deluxetable}{lcc}
\tablecaption{Systematic Uncertainties}
\tabletypesize{\scriptsize}
\tablewidth{0.8\textwidth}
\tablehead{\colhead{Type} & \colhead{Variation of Input Parameters} & \colhead{Impact on Results} \\ 
\colhead{} & \colhead{} & \colhead{} } 

%% All data must appear between the \startdata and \enddata commands
\startdata
Spectral bins & $\pm50\%$ & $<5\%$ \\
Spatial bins & $\pm50\%$ & $<1\%$ \\
Spatial template bins & $\pm50\%$ & $<1\%$ \\
Small ROIs & $+25\%$ & $\sim3\%$ \\
Number of free sources & $4\deg\rightarrow \theta_{200}+1\deg$\tablenotemark{a} & $<10\%$ \\
IRF uncertainties: Effective Area & $\pm10\%$\tablenotemark{b} & $<7\%$ \\
IRF uncertainties: PSF & $\pm15\%$\tablenotemark{b} & $<4\%$ \\
Diffuse model uncertainties & alternative diffuse models\tablenotemark{c} & 15-25\% more stringent limits \\
\enddata
\tablenotetext{a}{We chose a radius of 4\deg around each cluster center to account for photon contamination due to the PSF at low energies. In this test we modify the radius in which we leave the normalization free to vary within $\theta_{200}+1\deg$.}
\tablenotetext{b}{We employ the \textit{bracketing IRF} approach as discussed in \citet{Instrumentpaper} and use the tabulated values to scale the relevant IRF components.}
\tablenotetext{c}{We use a set of alternative diffuse emission models and replace the standard emission template used in the baseline analysis with these}
\tablecomments{Overview of systematic uncertainties as discussed in Section~\ref{sec:systematics}. We note that the largest impact on the results is due to the model for the Galactic diffuse emission.}
\label{tab:uncertainties}

\end{deluxetable*}
%\end{deluxetable}

\section{Summary and Conclusions}
\label{sec:summary}
We have used a dataset of 4 years of all-sky data from the \Fermi-LAT detector and performed a search for high-energy \gr\ emission originating from 50 X-ray luminous galaxy clusters. We specifically consider hadronically induced \grs\ originating from the ICM as described by the universal CR model by \citet{Pinzke2010} and employ a joint likelihood analysis to constrain the normalization of a common scale factor among clusters that is theoretically expected to describe the \gr\ luminosity of the ICM. In order to allow for different emission scenarios we categorize clusters in our sample by their morphologies and separately consider CC and NCC subsamples. 

We find evidence for excess emission at a significance of 2.7\,\ssigma, {naively taken as a first indication of \gr emission from galaxy clusters.} However, upon closer investigation we find that this global significance originates mainly from individual excesses present in three galaxy clusters: Abell~1367, Abell~3112 and Abell~400, the latter yielding a post-trial significance of 6.7\,\ssigma alone. For these three clusters, the LAT data alone cannot conclusively support or reject the hypothesis that the excess emission arises from within the clusters. The best-fit location of each excess is located within the virial radius of the respective cluster, but is offset from the cluster center.

With respect to the universal cosmic ray model we also note that the associated scale factors are significantly larger than the ones derived from other clusters in the sample. We also argue that in all three clusters there are individual radio galaxies which may be the origin of observed excesses. 

We establish bounds on the common scale factor $A_\gamma$, and use this to derive individual upper limits on the \gr\ flux. In addition, we use the jointly derived limit on $A_\gamma$ to calculate limits on the volume-averaged CR-to-thermal pressure $\langle X_\CR\rangle$. We compute median upper limits calculated within $R_{200}$, with the most stringent one being $\langle X_{\CR}\rangle<0.012$ for the combined sample and $\langle X_{\CR}\rangle<0.013$ and $\langle X_{\CR}\rangle<0.014$ for the CC and NCC sub-samples, respectively. { Assuming a linear dependence, our limits on $A_\gamma$ translate into a combined limit of the hadronic injection efficiency, $\zeta_{\rmn{p,inj}}$, by large scale structure formation shocks in the Mach number range $\mathcal{M}\simeq 3$--$4$ to be below 21\% for the combined sample and 25\% and 24\% for the CC and NCC clusters, respectively.} Removing the aforementioned three clusters that exhibit excess emission provides even more stringent limits on $A_\gamma$, which, for the combined sample yields $A_\gamma^{UL}=0.29$ that translates into $\langle X_{\CR}\rangle<0.008$ within $R_{200}$ and $\zeta_{\rmn{p,inj}}(\mathcal{M}\simeq3$--$4)<15\%$. Our limits on $\langle X_\CR\rangle$ and $\zeta_{\rmn{p,inj}}$ are the most stringent to date, {constraining hadronic} emission scenarios that predict astrophysical \grs\ originating in the ICM of galaxy clusters.\footnote{Note that all these limits are computed at the 95\% C.L.}

The systematic uncertainty associated with the modeling of the Galactic foreground emission represents the largest source of uncertainty affecting limits on extended emission from the ICM presented in this work. To account for this, we have tested our results against a set of alternative diffuse models spanning a range of interstellar emission model parameters. We find that the alternative models provide limits that differ from the baseline analysis by 20-30\%.

\begin{acknowledgments}
We thank the referee for a thoughtful report that helped improving the paper. The \Fermi-LAT Collaboration acknowledges generous ongoing support from a number of agencies and institutes that have supported both the development and the operation of the LAT as well as scientific data analysis. These include the National Aeronautics and Space Administration and the Department of Energy in the United States, the Commissariat \`a l'Energie Atomique and the Centre National de la Recherche Scientifique / Institut National de Physique Nucl\'eaire et de Physique des Particules in France, the Agenzia Spaziale Italiana and the Istituto Nazionale di Fisica Nucleare in Italy, the Ministry of Education, Culture, Sports, Science and Technology (MEXT), High Energy Accelerator Research Organization (KEK) and Japan Aerospace Exploration Agency (JAXA) in Japan, and the K.~A.~Wallenberg Foundation, the Swedish Research Council and the Swedish National Space Board in Sweden. 
Additional support for science analysis during the operations phase is gratefully acknowledged from the Istituto Nazionale di Astrofisica in Italy and the Centre National d'\'Etudes Spatiales in France. J.C. is a Wallenberg Academy Fellow. C.P. gratefully acknowledges financial support of the Klaus Tschira Foundation. AP acknowledges the NASA grant NNX12AG73G for support. {We thank our referee for the useful comments that improved the quality of this paper.} This research has made use of the NASA/IPAC Extragalactic Database (NED) which is operated by the Jet Propulsion Laboratory, California Institute of Technology, under contract with the National Aeronautics and Space Administration. 
\end{acknowledgments}

\bigskip % extra skip inserted% Create the reference section using BibTeX:
\bibliography{apj-jour,biblio_3authors}

%\begin{thebibliography}{9}   % Use for  1-9  references
%\begin{thebibliography}{99} % Use for 10-99 references
%\end{thebibliography}

{\it Facilities:} \facility{Fermi}



\appendix
\section{MC Simulation Studies}
\label{sec:validationsimulations}
We use the simulation package {\tt gtobssim} to efficiently generate MC realizations of the \gr\ sky using the parametrized instrument response. 

\subsection{Minimum energy threshold}
\label{sec:sim_emin}
The joint likelihood approach discussed in Section~\ref{sec:llh} makes the assumption that the individual data samples are uncorrelated. Assuming two sources $s_{1}$ and $s_{2}$, in the uncorrelated case, the composite $p$-value is $p_{\Sigma}=p_{1}\times p_{2}$, where $p_{1}$ and $p_{2}$ are the $p$-values associated with $s_{1}$ and $s_{2}$, respectively. This case corresponds to two sources that are far away from each another. In this case the difference between $p_{\rmn{joint}}$, which is the derived $p$-values from the joint likelihood, should be minimal, while for small distances, correlations impact the derivation of $p_{\rmn{joint}}$, and thus lead to an overestimation of the significance associated with this $p$-value. We assess this bias using an isotropic simulation with two identical power-law sources with $\Gamma=-2.3$ that were each modeled as an extended source with a disk of 2\deg in diameter. In Figure~\ref{fig:appdx_overlaps} we show the difference between $p_{\rmn{joint}}$ and $p_\Sigma$. We find that at small distances ($<3\deg$), for all minimum energy thresholds, there is a substantial bias due to overlaps. Towards larger distances, this bias is reduced. In this toy model, 3\deg corresponds to $R_{200}+1\deg$. However, by requiring larger distances between sources, we reduce the number of viable cluster candidates for our search. Hence we decided to use $E_{min}=500\mev$ as this threshold maximizes the number of clusters to be included (and thus the expected signal) while minimizing the bias on $p_{\rmn{joint}}$.

\begin{figure}[tbp]
\epsscale{.60}
\plotone{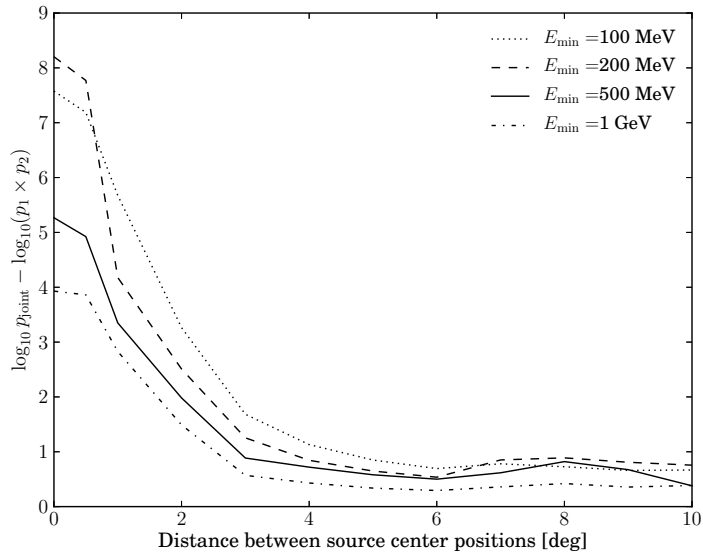}
\caption{\label{fig:appdx_overlaps} We show the difference between the jointly derived $p$-value and the case of uncorrelated samples, where $p_\Sigma=p_{1}\times p_{2}$ (see text for details) simulating two power-law sources with identical spectral model and modeled as a disk with 2\deg in diameter for a minimum energy threshold $E_{\rmn{min}}$ of (100\mev, 200\mev, 500\mev, and 1\gev). The solid line corresponds to our analysis threshold, chosen to minimize this overlap bias while maximizing the expected \gr flux.}
\end{figure}

\subsection{Significance assessment}

\begin{figure}[tbp]
%\epsscale{.95}
\plotone{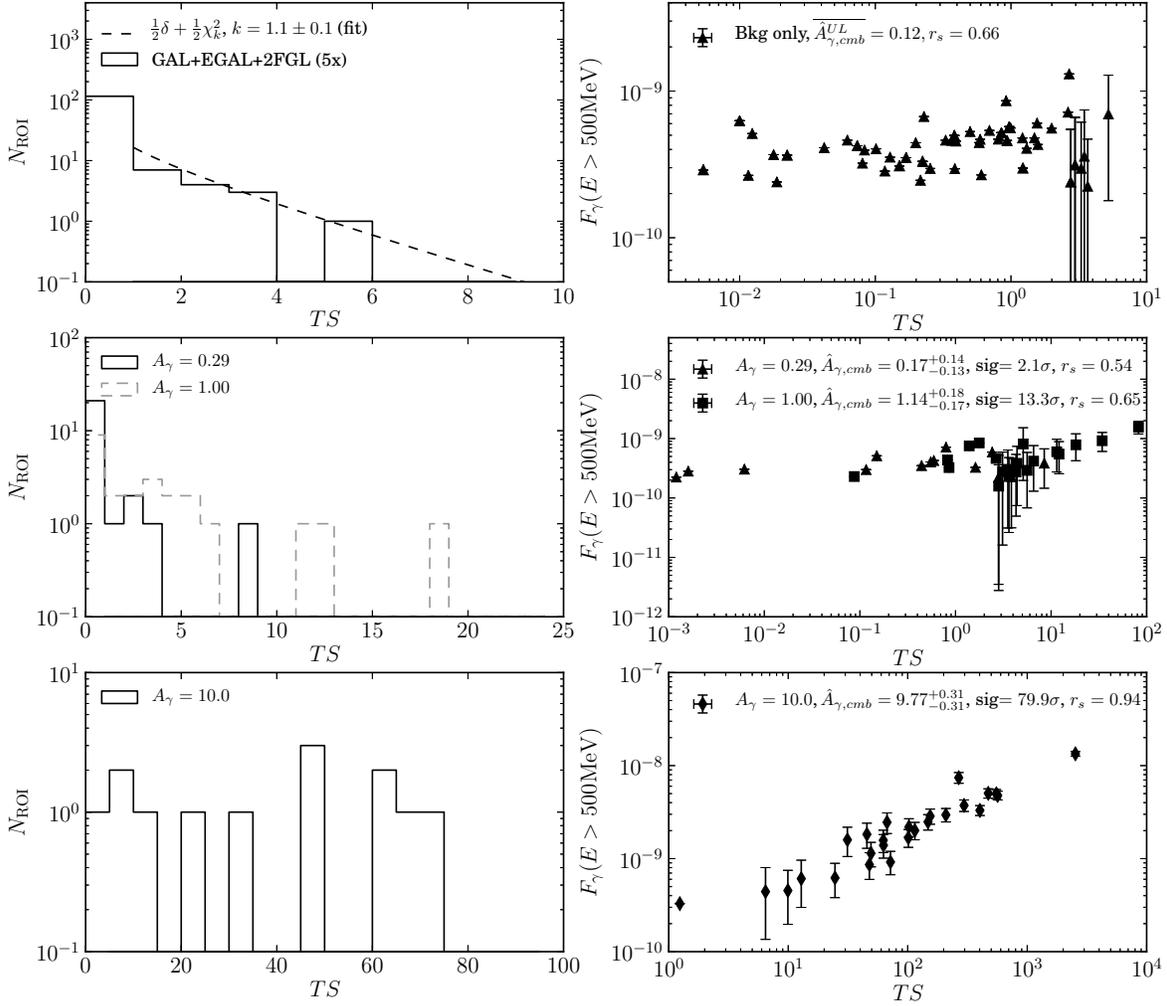}
\caption{\label{fig:appdx_tsdistrib} In the top left panel we show the $TS$-distribution for five background-only simulations and fit the distribution to the null hypothesis according to our analysis (refer to Section~\ref{sec:llh} for details). In the top right panel we show the constraints on the \gr\ flux using the ROI-specific individual scale factors $A_{\gamma,i}$ and relate them to the associated $TS$ values. $r_{s}$ denotes the Spearman rank correlation coefficient and \emph{sig} refers to the combined significance derived from $TS_{\rmn{global}}$. The middle and lower panel show the same but assuming a putative \gr\ source in addition to the background. We show that for $A_\gamma=1$ and even more so for $A_\gamma=10$ the $TS$-distributions clearly depart from a $\chi^2$ distribution. It should be noted however, that while the background simulation is based on five MC realizations of the \gr\ sky, the various signal simulations represent a single example realization for each assumed signal.}
\end{figure}

\label{sec:sim_significance}
In Figure~\ref{fig:appdx_tsdistrib} we show ROI-specific $TS$-distributions for a background only simulation (top). We find that the $TS$-distribution in the background-only case can be well described by $\frac{1}{2}\delta+\frac{1}{2}\chi_{k}^{2}$ and obtain $k=1.1\pm0.1$ in an unbinned maximum likelihood fit, giving rise to the usual definition of significance as $\sqrt{TS}$. 

\subsection{Signal studies}
In addition, we include the results from simulations including a (weak) putative CR-induced \gr\ signal corresponding to $A_\gamma=0.29$ (nominal best-fit value for the combined sample) as well as $A_\gamma=1$ (middle panel in Figure~\ref{fig:appdx_tsdistrib}), i.e., assuming the predictions by \citet{Pinzke2011}. Finally, we repeat the analysis assuming a strong signal, characterized by $A_\gamma=10$ (bottom panel in Figure~\ref{fig:appdx_tsdistrib}). For all simulations the true value of $A_\gamma$ is recovered in the combined result, validating our analysis approach, although the global significance varies with respect to the signal simulation. For the background-only case, only upper limits can be derived. Assuming the nominal best-fit value for $A_\gamma$ from the combined sample, yields a combined significance of 2.1\,\ssigma. For $A_\gamma=1$ and $A_\gamma=10$ these values are much higher yielding a global $TS$ value corresponding to a significance of 13.3\,\ssigma and 79.9\,\ssigma, respectively. 

Given that a simulation with $A_\gamma=0.29$ yields a combined significance comparable with what we have found in our analysis, we investigated how this signal could be studied further. To that end we compare the expected \gr\ flux based on the ROI-specific scale factors $A_\gamma,i$ with their associated ROI $TS$ values and quantify the correlation through calculating Spearman-rank correlation coefficient \citep{Spearman1904}, denoted by $r_{s}$. We find Spearman-rank coefficients $>0.5$ indicating a correlation. However, even for the background case, $r_{s}=0.6$ is obtained which is the same as what we find in the signal case for $A_\gamma=1$. This illustrates that a correlation analysis with such a weak signal, further studying the excess we find, is difficult. Only for the strong signal of $A_\gamma=10$ we find a correlation coefficient $r_{s}=0.94$ indicating a strong correlation between expected \gr flux and associated ROI $TS$ value. 

\section{Analytic Cosmic Ray Model}
\label{sec:CRs}

Following the analytic CR formalism
\citep{2004A&A...413...17P,2008MNRAS.385.1211P,Pinzke2010}, we obtain the volume-weighted, energy-integrated, and omnidirectional (i.e., integrated over the $4\pi$ solid angle) \gr\ source function
due to pion decay, 
\begin{equation}\label{eq:gammaray_analytic}
  \lambda_{\pig}(R,E) = \kappa_{\pig}(R)\,\lambda_{\pig}(>E)\,.
\end{equation}
Here the spatial part of the \gr\ emission is determined by
\begin{equation}
  \label{eq:gammaray_analytic_spatial}
  \kappa_{\pig}(R) = C_\CR(R)\,\rho_\gas(R)\,,
\end{equation}
where the CR proton distribution is
\begin{eqnarray}
  \label{eq:tildeCM}
\frac{\tilde{C}_\CR(R)}{\rho_\gas(R)} &=& \left(C_{\rm 200}- C_{\rm center}\right)\left(1 +
\left(\frac{R}{R_{\rm trans}}\right)^{-\beta}\right)^{-1} + C_{\rm
 center},  ~\rmn{and}\\
  \label{eq:C-scaling1}
  C_{\rm 200}   &=& 1.7\times 10^{-7}~~\times(M_{200} / 10^{15}\,M_\odot)^{0.51}, \\
  \label{eq:C-scaling2}
  R_{\rm trans} &=& 0.021\,R_\rmn{200}~~\times(M_{200} / 10^{15}\,M_\odot)^{0.39}, \\
  \label{eq:C-scaling3}
  \beta       &=& 1.04~~\times(M_{200}/10^{15}\,M_\odot)^{0.15},
\end{eqnarray}
where $\tilde{C} = C m_p/\rho$ denotes the dimensionless normalization
of the CR distribution function.  For massive clusters ($M_{200} \sim
10^{15}\,M_\odot$) the CR distribution traces the gas density, while the CR
density is slightly enhanced in the center for smaller systems. Note,
however, that the \gr\ flux depends only weakly on the exact CR
density in the center (i.e., $C_{\rm center}$) since most of the flux
originates from outside the transition region.

The spectral part of Eq.~\eqref{eq:gammaray_analytic} for the
photon energies relevant to \Fermi-LAT ($100\mev\lesssim E_\gamma \lesssim
1\,\rmn{T}e\rmn{V}$) is given by
\begin{eqnarray}
\lambda_{\pig}(>E) &=& \frac{4m_{\pi^0} c}{3m_\p^3}
  \times \sum_{i=1}^3\frac{\sigma_{\rmn{pp},\,i}}{\alpha_i\,\delta_i}
\left(\frac{m_\p}{2m_{\pi^0}}\right)^{\alpha_i}
\Delta_i\left[\mathcal{B}_x\left(\frac{\alpha_i+1}{2\delta_i},
\frac{\alpha_i-1}{2\delta_i}\right)\right]_ {x_1}^{x_2} \, ,\nonumber\\
\label{eq:gammaray_analytic_energy}
& &\rmn{and} \qquad  
x_j=\left[1+\left(\frac{m_{\pi^0}c^2}{2E_{\gamma,j}}\right) ^{2\delta_i}\right]^{-1}\,,
\end{eqnarray}
where the sum over $i$ extends over $\boldsymbol{\Delta}=(0.767,0.143,
0.0975)\,,~\rmn{and}~~\boldsymbol{\alpha}=(2.55, 2.3, 2.15)$. 
The \gr spectrum rises until a maximum at the mass of the $\pi^0$ meson followed by a concave shaped tail determined by the universal CR spectrum. The shape parameter $\delta_i \simeq 0.14\,\alpha_i^{-1.6}+0.44$ allows us to accurately predict the emission close to the pion bump in combination with the effective inelastic cross-section for proton-proton interactions, $\sigma_{\rmn{pp},\,i} \simeq 32\,(0.96+\e^{4.42-2.4\alpha_i})\, \rmn{mbarn}$. We have also have introduced the abbreviation 
\begin{equation}
\left[\mathcal{B}_x\left(a,b\right)\right]_{x_1}^{x_2}=
\mathcal{B}_{x_2}\left(a,b\right)-\mathcal{B}_{x_1}\left(a,b\right)\,,
\end{equation}
where $\mathcal{B}_x\left(a,b\right)$ denotes the incomplete
Beta-function.

\section{Relationship Between CR Acceleration Efficiency and Gamma-ray Flux}
\label{sec:acc_eff}

\begin{figure}[tbp]
\epsscale{.60}
\plotone{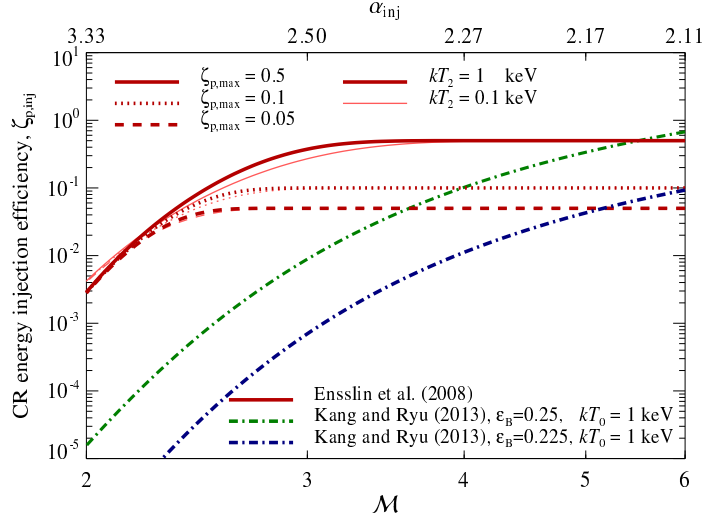}
\caption{{(color online) Acceleration efficiency $\zeta_{\rmn{p,inj}}$ as a
    function of shock Mach number ($\mathcal{M}$) for two different post-shock
    temperatures of 0.1~keV (thick bright) and 1~keV (thin faded). We show the
    acceleration efficiency that was used in our simulations ($\zeta_\rmn{p,max}
    = 0.5$, red solid) and scaled to different values for the saturated
    acceleration efficiency ($\zeta_\rmn{p,max} = 0.1$ and $\zeta_\rmn{p,max} =
    0.05$; red dotted and dashed, respectively). Also shown in blue and green is
    the acceleration efficiency in a model by \citet{2013ApJ...764...95K} for
    different values of $\varepsilon_B$, the ratio of downstream
    turbulent-to-background magnetic field, which determines the the injection
    efficiency in their model. Since shocks with Mach numbers
    $\mathcal{M}\simeq3$--$4$ are mostly responsible for injecting CRs in
    clusters, constraints on $\zeta_{\rmn{p,inj}}$ in the model by
    \citet{2007A&A...473...41E} are more conservative in comparison to the model
    by \citet{2013ApJ...764...95K}.}}
\label{fig:zeta_max}
\end{figure}

{In this Appendix, we investigate how we can use upper limits on the
  $\gamma$-ray flux to constrain the CR injection efficiency in various models
  for CR acceleration. The CR model adopted in this paper \citep{Pinzke2010} is
  based on a simplified scheme \citep{2007A&A...473...41E} to compute the
  CR-energy acceleration efficiency at shocks (in units of the shock-dissipated
  thermal energy, corrected for adiabatic compression),
  $\zeta_\rmn{p,inj}(\mathcal{M}) = \varepsilon_\CR/
  \varepsilon_{\mathrm{diss}}$. It employs the thermal leakage model
  \cite[e.g.,][]{1984ApJ...286..691E, 1994APh.....2..215B, 1995ApJ...447..944K},
  which conjectures a momentum threshold for injection that is a constant
  multiple ($x_\rmn{inj}=3.5$) of the peak thermal momentum, at which the CR
  power-law distribution connects to the post-shock Maxwellian.  More refined
  models \cite[such as in][]{2011ApJ...734...18K, 2013ApJ...764...95K} are
  motivated by non-linear shock acceleration and fix the injection momentum by
  the consideration that the particle speed should be several times larger than
  the downstream flow speed in order for suprathermal particles to diffuse
  upstream across the shock transition layer.  This yields a Mach-number
  dependent $x_\rmn{inj}$, which increases for weaker shocks such that a
  progressively smaller fraction of particles can participate in the process of
  diffusive shock acceleration.  As a result, for the preferred values of the
  ratio of downstream turbulent-to-background field, $\varepsilon_B\gtrsim0.25$,
  those models predict $\zeta_\rmn{p,max} \gtrsim 0.4$--$0.5$, depending on the
  existence of a pre-existing CR population.\footnote{This is obtained by taking
    the ratio of CR acceleration efficiency ($\eta$) and gas thermalization
    efficiency ($\delta$) in the limit of large Mach numbers
    \citep{2013ApJ...764...95K}.} While those values for the acceleration
  efficiency are now challenged by $\gamma$-ray observations of supernova
  remnants, additional physics (such as amplification mechanisms of the magnetic
  field) may lower the value of $\varepsilon_B$ and cause the acceleration
  efficiency to saturate at lower values \citep{2013ApJ...764...95K}.  Moreover,
  in weak heliospheric shocks additional shock phenomena (e.g., whistler waves
  in the shock front, etc.) are observed that may add to the uncertainty of the
  acceleration efficiency. In summary, while the models for the acceleration
  efficiency in shocks became more detailed and physical over the last year, new
  observations point to the necessity of further improving those models.}

{In Figure~\ref{fig:zeta_max} we show the relation between acceleration
efficiency and shock Mach number. For the relevant energy regime that we
consider in this work ($E_\gamma\gtrsim 500\mev$), our CR model predicts a CR
spectral index of $\simeq 2.4$ that flattens toward higher energies
\citep{Pinzke2010}. This implies that shocks with Mach numbers $\mathcal{M}
\gtrsim 3.5$ (depending somewhat on the post-shock temperature) are the most
relevant for the CR budget. Because the acceleration efficiency has already
saturated for this range of shock strengths in the model of
\citet{2007A&A...473...41E}, the CR pressure and thus the hadronically induced
\gr luminosity are approximately proportional to $\zeta_\rmn{p,max}$.  For
different acceleration models \citep[such as in][]{2013ApJ...764...95K}, these
upper limits provide interesting constraints on the acceleration efficiency at
intermediate strength shocks of Mach number $\mathcal{M}\simeq3$--$4$, a regime
complementary to that studied at supernova remnant shocks.}

\section{Radio and X-ray Sources in the Field of View of A3112, A1367 and A400}
\label{sec:mwdiscussion}
\begin{figure*}
\begin{center}
\includegraphics[width=\columnwidth]{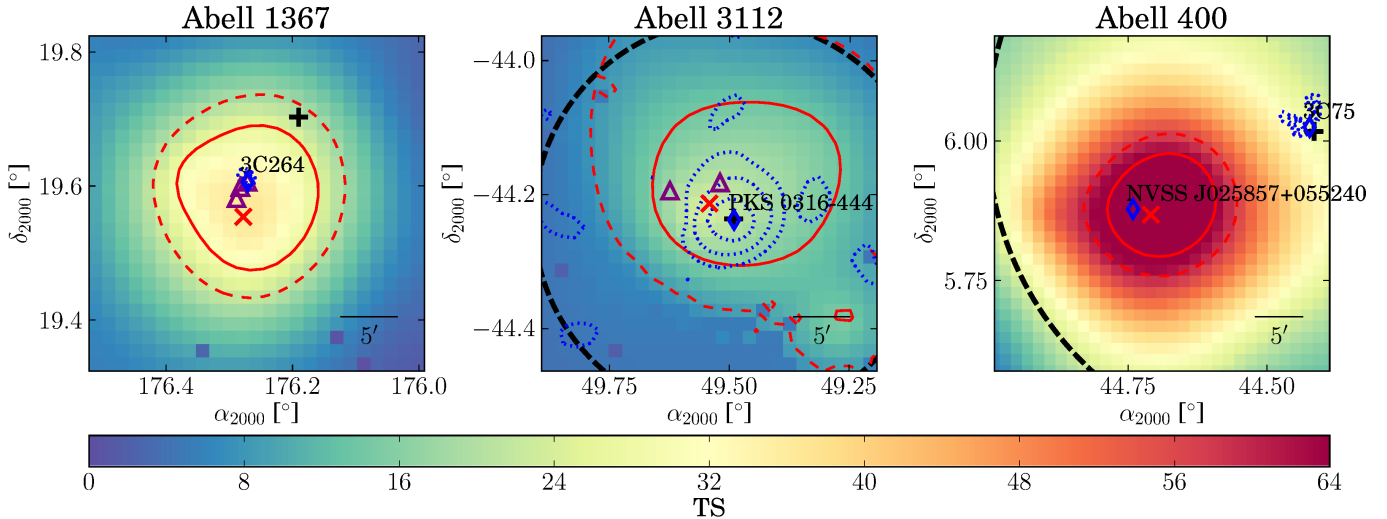}
\end{center}
\caption{\label{fig:TSMW}(color online) {$TS$ maps of our three cluster candidates with $TS>9$. Each map shows a $0.5\deg\times0.5\deg$ section ($0.6\deg\times0.6\deg$ for Abell~400) that was recentered to the best-fit position obtained using {\tt{gtfindsrc}}. The best-fit position from the refined $TS$ calculations is marked as red $\times$. Shown in red are the $3\,\sigma$ (solid) and $4\,\sigma$ contours (dashed). The blue diamond-shaped points correspond to the NED positions of the radio galaxies as discussed in the text. The purple upright open triangles denote Chandra X-ray sources that fall within the error circle of the \gr point source. Overlaid radio contours (blue, dotted) were obtained from the NVSS for A1367 and A400 \citep{Condon1998}. The radio contours for A3112 were obtained from the Parkes-MIT-NRAO Survey \citep{Condon1994}. The cluster center is marked by a black cross in each panel. The virial radius of the cluster is indicated by the dashed black line (partially visible in the maps for both Abell~3112 and Abell~400). Coordinates taken from NED/SIMBAD. Each pixel is 0.2\deg across.}}
\end{figure*}
{
In this section we provide a discussion of the radio sources in the field of view of the three clusters as discussed in Section~\ref{sec:exceptions}. We note that a detailed characterization of the origin of the excess emission is beyond the scope of this work. Based on the refined best-fit positions from our higher resolution TS-map, we performed a search for sources within the $3\,\sigma$ contours as shown in Fig.~\ref{fig:TSMW}. Below we provide supplemental information regarding the three clusters discussed in the main text.}
\begin{itemize}
\item{{For Abell 1367, our best-fit position is consistent with that of the radio galaxy \object{3C264} \citep{Fey2004}. As we cannot distinguish between Abell~1367 and 3C264, we conclude by similarity with previous \gr detections in clusters that we likely observe \gr emission from the radio galaxy (e.g., M87 in Virgo or IC310 in Perseus).}}
\item{{Similarly, the origin of the emission towards Abell~3112 may be from the radio galaxy \object{PKS 0316-444}, which is located $4.2'$ away from the best-fit position of the excess \citep{Costa1998}.}}
\item{{In the vicinity of the best-fit position of the excess towards A400, a (not further classified) radio source \object{NVSS J025857+055240} was reported by \citet{Condon1998}. Due to this positional coincidence, it is plausible to attribute the observed \gr emission to this object, although this hypothesis warrants further investigation.}}
\end{itemize}
{
In addition to the previously discussed radio sources, there are multiple Chandra X-ray sources that fall within the error circle of the best-fit positions for the excesses in Abell~1367 and Abell~3112 respectively. While the association of the excesses in \grs with individual radio galaxies is well in line with previous \gr detections, e.g., M87 in Virgo or NGC1275 and IC310 in Perseus, the similarity argument we present here is not sufficient to claim detection of \grs from these respective objects. }

\section{ROI-specific Counts/Model Comparison}
\label{sec:spectra}
We provide for each ROI the observed photon counts in each energy bin along with the predicted model counts from the best-fit background-only model and show this in Figures~\ref{fig:spectra}--\ref{fig:spectra3}. 

\begin{figure*}[tbp]
%\epsscale{.99}
%\plottwo{spectra_pad1.eps}{spectra_pad1.eps}
\plotone{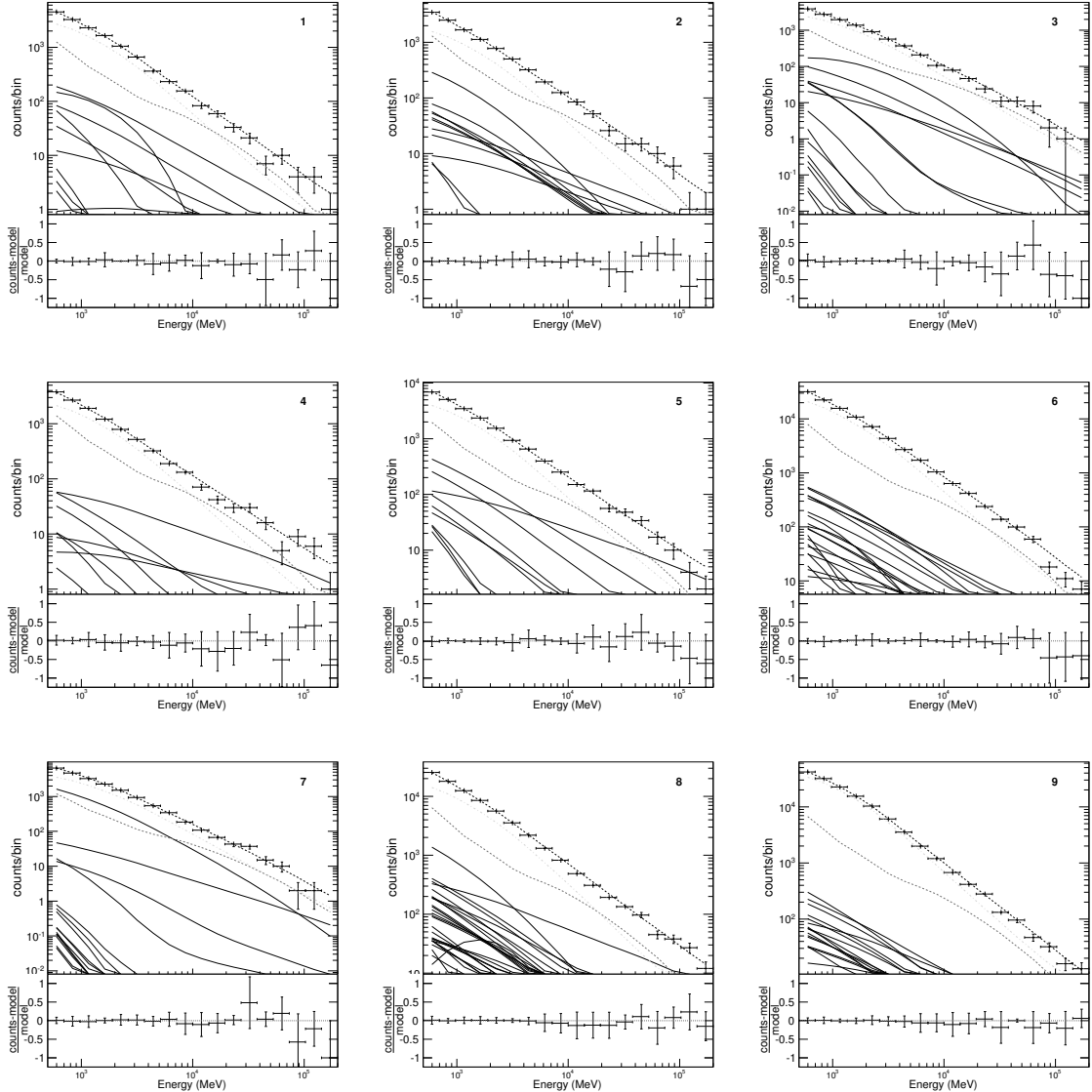}
\caption{\label{fig:spectra}Fitted counts spectra for 9 of the 26 analysis ROIs without additional cluster sources. The crosses indicate the measured counts in each energy bin while the black dashed lines show the total sum of all model counts for all components. The gray-dashed lines refer to the Galactic diffuse component while the gray-dotted lines correspond to the isotropic extragalactic diffuse component. Solid lines indicate additional background sources. We obtain reasonable fits in all energy bins. The lower panel for each ROI shows fractional residual counts integrated over the entire ROI.}
\end{figure*}

\begin{figure*}[tbp]
%\epsscale{.99}
%\plottwo{spectra_pad2.eps}{spectra_pad2.eps}
\plotone{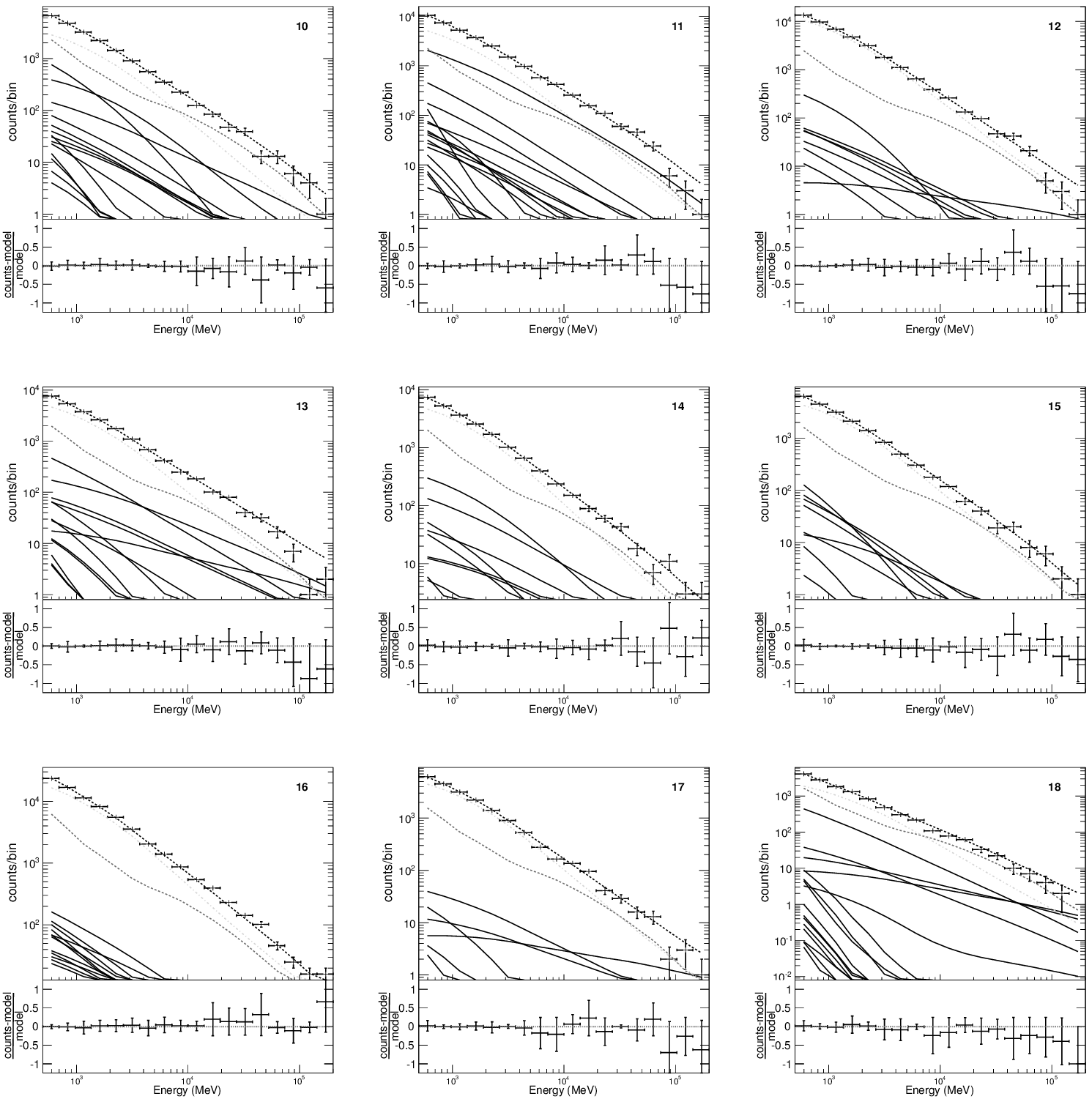}
\caption{\label{fig:spectra2} Fitted spectra for ROIs 10-18. See caption of Figure~\ref{fig:spectra} for details.}
\end{figure*}

\begin{figure*}[tbp]
%\epsscale{.99}
%\plottwo{spectra_pad3.eps}{spectra_pad3.eps}
\plotone{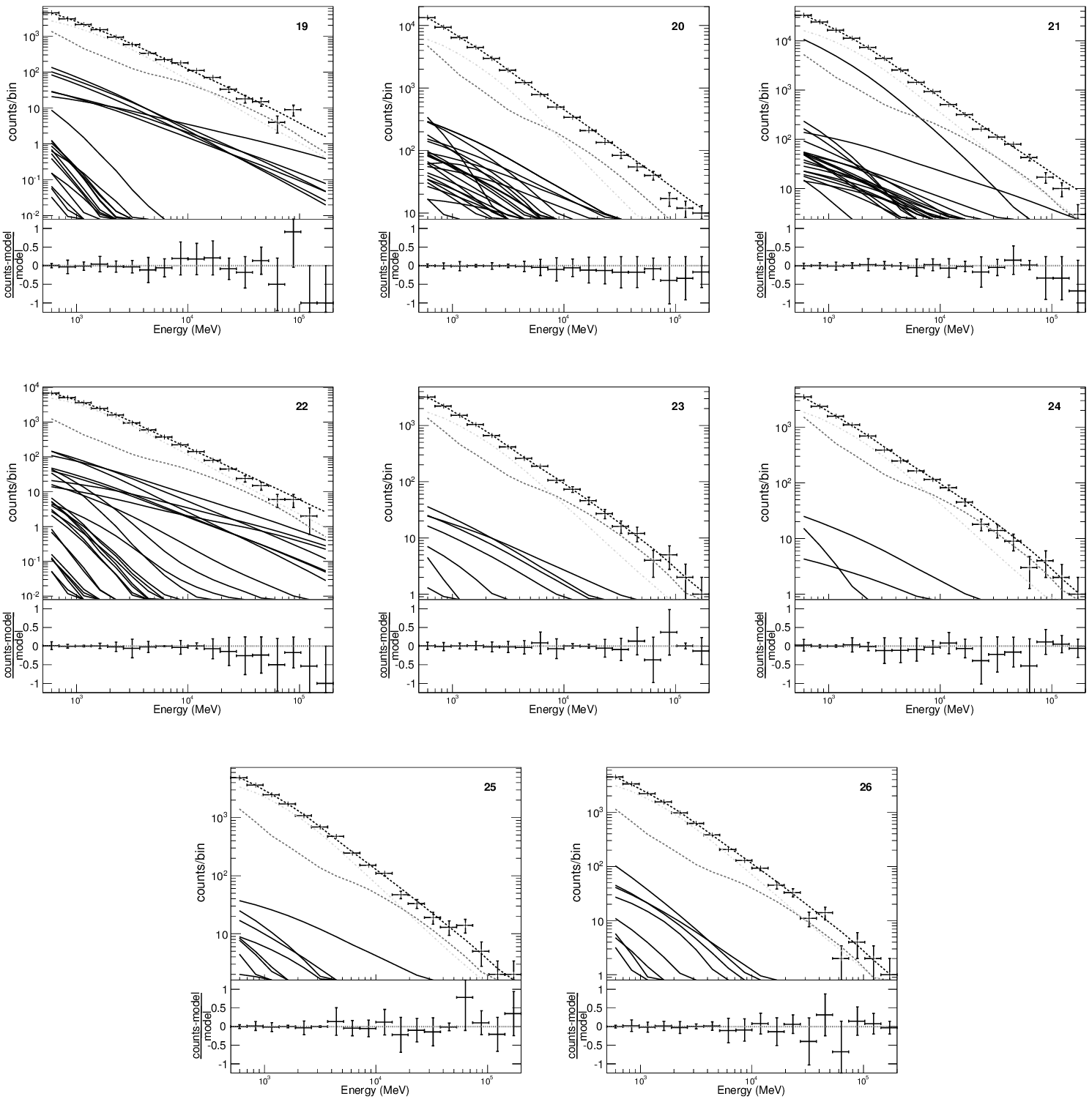}
\caption{\label{fig:spectra3} Fitted spectra for ROIs 19-26. See caption of Figure~\ref{fig:spectra} for details.}
\end{figure*}

%% \clearpage
%% \LongTables
%% \begin{landscape}
%% \input{tab1.tex}
%% \end{landscape}

%% \clearpage
%% \LongTables
%% \begin{landscape}
%% \input{tab4.tex}
%% \end{landscape}

\end{document}